\documentclass[10pt,a4]{article}
\usepackage[a4paper,bottom=1in,top=1in,left=1in,right=1in]{geometry}
\usepackage{graphicx}
\usepackage{times} 
\usepackage{graphicx} 
\usepackage{dcolumn}
\usepackage{bm}
\usepackage{amssymb}
\usepackage{amsmath}
\usepackage{wasysym}
\usepackage{color}
\usepackage{hyperref}
\usepackage[normalem]{ulem}
\usepackage{float}
\usepackage{flushend}
\usepackage{multirow}
\usepackage{cancel}
\usepackage{cite}
\usepackage[dvipsnames]{xcolor}
\hypersetup{
colorlinks,
citecolor=blue,
filecolor=blue,
linkcolor=blue,
urlcolor=blue}

\usepackage{MnSymbol}
\usepackage{xcolor}
\definecolor{C3}{HTML}{DC143C}
\definecolor{C2}{HTML}{458B00}
\definecolor{C0}{HTML}{1874CD}
\definecolor{C4}{HTML}{1874CD}
\definecolor{C5}{HTML}{1874CD}
\definecolor{C6}{HTML}{1874CD}
\newcommand{\citep}[1]{\cite{#1}}
\newcommand{\citet}[1]{\cite{#1}}

\title{Mixed baroclinic convection in a cavity}

\author{Abhishek Kumar$^{1,}$\thanks{Email: abhishek.kir@gmail.com}\,\, and  Alban Poth{\'e}rat$^{1}$\\
 Centre for Fluid and Complex Systems, Coventry University, Coventry CV1 5FB, United Kingdom}

\date{}

\begin{document}

\maketitle

\abstract{We study the convective patterns that arise in a nearly { semi-cylindrical} cavity fed in with hot fluid at the upper 
boundary, bounded by a cold, porous semi-circular boundary at the bottom, and infinitely extended in the 
third direction. While this configuration is relevant to continuous casting processes that are significantly more complex, we focus on the flow patterns associated with the particular form of mixed 
convection that arises in it. Linear stability analysis and direct numerical simulations (DNS) are conducted, 
using the spectral element method to identify observable states. 
The nature of the bifurcations is determined through Stuart--Landau analysis for completeness.
The base flow consists of two counter-rotating rolls driven by the baroclinic imbalance due 
to the curved isothermal boundary. These are however suppressed by the through-flow, which is 
found to have a stabilising influence as soon as the Reynolds number $Re$ based on the through-flow 
exceeds 25.
For a sufficiently high Rayleigh number, this base flow is linearly unstable to three different modes, 
depending on $Re$. For $Re\leq75$, the rolls destabilise 
through a supercritical bifurcation into a travelling wave. For $100\leq Re \leq 110$, a subcritical bifurcation leads to a standing oscillatory 
mode, whereas for $Re\geq150$, the unstable mode is non-oscillatory and grows out of a 
supercritical bifurcation.
The direct numerical simulations confirm that 
in all cases, the dominant mode returned by the linear stability analysis precisely matches the 
topology and evolution of the flow patterns that arise out 
of the fully nonlinear dynamics.}

\section{Introduction}
This work is concerned with the convective patterns arising in cavities with curved isothermal boundaries and permeated by a through-flow. This configuration is typical of continuous casting processes of 
metallic alloys. In this type of process, solidified metal is pulled from 
the bottom of a pool of melted metal continuously fed from above. The pulling speed is adjusted to match 
that of the solidification front which, therefore, behaves as a steady but porous boundary for the fluid.
Problems of this class, and solidification problems in general, are governed by a complex interplay 
of coupled phenomena, among which: thermal and chemical convection-diffusion involving multiple species, 
hydrodynamic instabilities arising in boundary layers and shear regions, solidification and the 
solid-liquid phase interaction that ensues both at the boundaries (in mushy layers) and the bulk 
(transport of solid particles). 
A complete description of this process requires extensive modelling of how these 
phenomena are coupled, usually at the expense of strong modelling assumptions (see for example 
\cite{kuznetsov1997_cht, sheng2000_mmtb, brian2001_isiji}). A key mechanism among these involves 
the rather unusual 
type of mixed convection coupling the buoyancy indirectly caused by the shape of the boundaries and the 
through-flow. Instead of attempting a full description of the industrial process, we focus on 
the physical process associated to this particular coupling.

\begin{figure}
\begin{center}
\includegraphics[scale = 1,trim={0 1cm 0 0},clip]{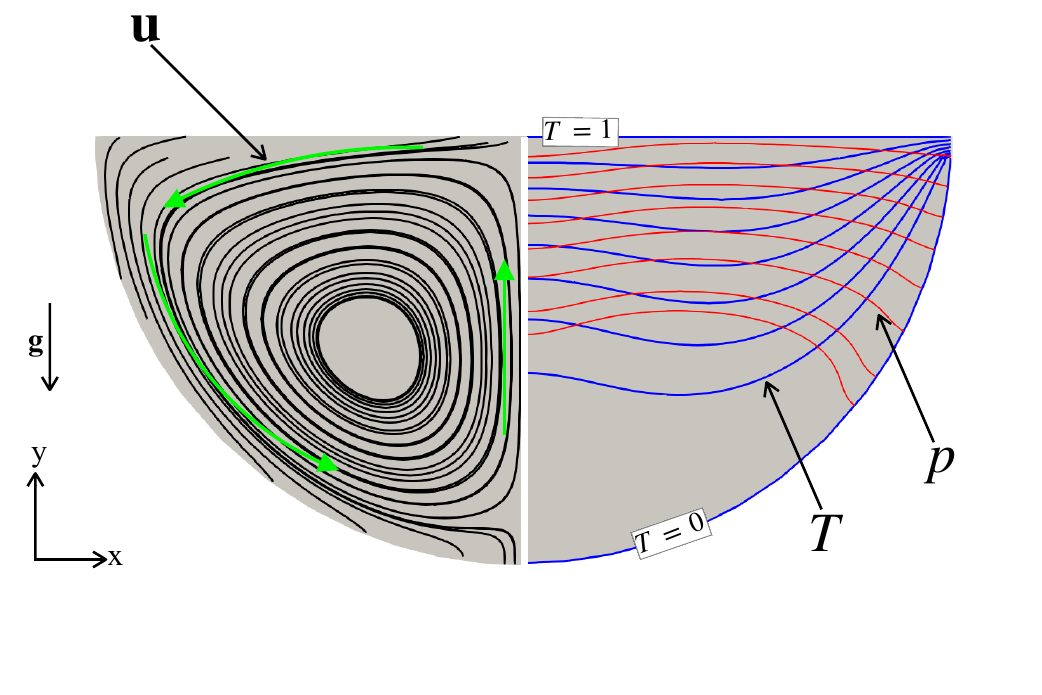}
\end{center}
\caption{Flow configuration calculated at $Re=25$ and $Ra=10^4$. Left: streamlines (black), right: isobars (red) and isotherms (blue). While the convection is initiated by the baroclinic imbalance 
near the corners, where isobars and isotherms intersect, the rolls result from the return flow generated 
in the middle, where the jets initiated opposite corners meet.
\label{fig:flow_configuration}}
\end{figure}

The first physical ingredient in it is the source of buoyancy. Unlike configurations of the 
Rayleigh--B\'enard type, the stratification itself is not a direct source of instability, as the hot - hence 
lighter - fluid is fed at the top surface of the pool, while the cold fluid is located at the bottom 
near the solidification front. Instead, the convection originates from the shape of the isothermal solidification front. { These intersect isobars that are mostly horizontal in the 
bulk but curved near the boundaries (see figure \ref{fig:flow_configuration})}. Consequently, the pressure forces cannot oppose the fall of 
heavy fluid along the boundary and a fluid motion must exist, no matter how small the temperature 
difference between the hot and the cold boundaries. Barotropic buoyancy sources of this kind are 
mostly studied in the context of oceans and atmospheres, where the 
misalignment of density and pressure gradients stems from the pressure contribution of the centrifugal 
forces due to planetary rotation \cite{hart1979_arfm,pierrehumbert1995_arfm}. Nevertheless, the simplest manifestation of baroclinic imbalance is obtained by tilting the plane configuration of the Rayleigh--B\'enard problem away from the horizontal position. In an infinite geometry, a flow along 
the tilted direction is driven by the temperature gradient. At low tilt angles, 
the base convective flow destabilises to transversal disturbances under the form of non-oscillating 
rolls at low Prandtl number and travelling waves at high Prandtl number (the Prandtl number $Pr=\nu/\alpha$ is the ratio of viscosity $\nu$ to thermal diffusivity $\alpha$). At higher tilts, longitudinal 
rolls dominate \citep{hart1971_jfm,korpela1974_ijhmt}. While the saturated state may involve nonlinear 
interaction between transverse and longitudinal { modes} if their respective critical Rayleigh numbers are 
close to each other, the travelling wave is by contrast always subject to a secondary instability 
and not observable \citep{fujimura1993_jfm}.

The second ingredient is the through-flow fed in at the upper boundary of the cavity and 
escaping at the lower boundary through solidification. Through-flows are found in two types of 
mixed convection problems of some relevance to our configuration: at the boundary of a heated cavity 
\citep{papanicolaou1992_jfm}, or through a heated conduit (see \cite{jaluria1980} and \cite{kelly1994_aam} for reviews), whether a pipe \citep{shome1995_ijhmt}, a duct \citep{nicolas2000_ijhmt} or a channel \citep{gage1968_jfm}. In all conduit configurations, the shear associated 
to the through-flow acts as a source of instability. {For example, in channels 
Tollmien--Schlichting waves are favoured in this way \citep{schmid01}} and the mixed convection patterns 
result from a competition between {buoyancy-driven and hydrodynamically-driven} instabilities. Whether one or the other dominates 
depends on the ratios of buoyancy, and inertia to viscous forces, respectively measured by the Rayleigh 
number $Ra= {\beta g \Delta T h^3}/({\nu \alpha})$ 
and the Reynolds number $Re=U_0h/\nu$ ($\beta$,  $\Delta T$, $h$, $g$, $U_0$ are the fluid's thermal expansion coefficient, a temperature difference between boundaries, a domain diameter, gravitational acceleration, and a fluid inlet velocity). In the simplest configuration of the Rayleigh--B\'enard--Poiseuille problem, 
transversal rolls dominate at low Reynolds numbers while Tollmien--Schlichting waves characteristic 
of the Poiseuille flow problem dominate for $Re>140$ \citep{fujimura1995_pf}. In rectangular cavities, 
by contrast, natural convection sets in through an oscillatory mode if the hot wall is located on the 
side \citep{briggs1985_jht}. The first unstable mode remains oscillatory when mixed convection is 
introduced, with a through-flow along the top wall. It sets in through a ``rolling pad"  instability for 
sufficiently high Richardson number $Gr/Re^2$, where the Reynolds number is based on the through-flow and $Gr=Ra Pr^{-1}$ is the Grashof number. Hence the through-flow tends to suppress convection \citep{papanicolaou1992_jfm}. 
This configuration bears important similarities with the problem we are considering, in that 
isotherms and isobars are not aligned and the thermal instability is damped by a 
through-flow. {In both configurations, the base temperature gradient induces a stable 
stratification, so convection does not ensue from a Rayleigh-B\'enard instability.} Nevertheless, the 
shape of the boundaries differs considerably between the two problems 
and the through-flow is only local in \cite{papanicolaou1992_jfm}'s work. 

Indeed, a specifically interesting aspect of the cavity configuration is that during continuous casting, the accumulating solid phase at the solidification front is continuously pulled downward, so the lower boundary remains at a constant position. This feature of the process is modelled by means of a porous boundary condition accounting for the mass flux from the liquid to the solid phase, as proposed by \cite{Flood:MST1994}. As 
such, it lacks the shear responsible for the hydrodynamic part of the instability  {in 
other mixed convection problems such as the Rayleigh--B\'enard--Poiseuille problem}, and mainly acts to 
suppress the base convective flow. The second specificity is that unlike most other problems of mixed 
convection, the source of buoyancy is purely baroclinic, {and not due to an unstable 
stratification (see figure~\ref{fig:flow_configuration} for the flow configuration)}. While a background shear can inhibit baroclinic instabilities in 
open flows \citep{james1987_jas}, the instabilities arising from the interaction of the uniform flow with baroclinic 
buoyancy in the confined fluid domain we are considering are not known.
Because of these specificities, the problem of a heated flow through a cavity may possess a very 
different phenomenology to that encountered in the problems involving mixed convection discussed above, 
even though they share some of their ingredients. As such, the
 convective patterns, whether they are steady or not and the nature of the bifurcation associated to their onset are not known, despite their central role in solidification problems.

The purpose of this work is precisely to identify both the mechanisms governing the stability of a generic 
flow supporting this phenomenology, and the actual flow that ensues. The minimal geometry 
with all the necessary ingredients for this purpose was first proposed by \cite{Flood:MST1994}. It consists 
of a pool with a hot, isothermal, rigid, free-slip upper boundary supporting a uniform inflow, and a 
cold isothermal semi-circular solid wall representing the solidification front at the bottom.  
The stable stratification avoids 
the complication of mechanisms associated to unstably stratified flows, even though these could occur in 
some solidification problems depending on the nature of the alloys being solidified \citep{kuznetsov1997_cht}. For 
simplicity, the domain shall be assumed infinitely extended in the third direction.
We tackle the problem numerically, with both Linear Stability Analysis (LSA)  and Direct Numerical Simulations (DNS)
based on a combination of the spectral element method and Fourier-spectral discretisation in the invariant direction. This choice of methodology offers the necessary flexibility to deal with the non-trivial shape of the boundary whilst retaining the numerical precision of spectral methods \citep{Canuto:book:SpectralFluid,karniadakis_spectral:book}. 
We shall seek answers to the following questions:
\begin{enumerate}
\item What is the nature of the base flow?
\item In which conditions is this flow stable?
\item What is the topology and nature (oscillatory or not) of the unstable mode?
\item What is the nature of the bifurcation at the onset of the instability?
\end{enumerate}

After the mathematical definition of the problem and governing equations
in \S~\ref{sec:problem_formulation}, we shall provide the details of the numerical methods we use and of their
 validation (\S~\ref{sec:comp}). We shall then determine the base flow by means of 
two-dimensional DNS in the vertical plane (\S~\ref{sec:2dsol}) and  assess its stability to infinitesimal 
 three-dimensional perturbations through linear stability analysis (\S~\ref{sec:lsa}). 
Three-dimensional DNS of the flow near the onset of stability shall provide a validation for the LSA 
approach, and indicate whether the saturated state can be inferred from it. The relevance of the 
LSA approach shall be further validated by seeking the nature of the bifurcation by means of a 
Stuart--Landau analysis \citep{Sheard:JFM2004} in \S~\ref{sec:ST_model}. Finally, concluding remarks are presented in \S~\ref{sec:conclusion}.

\section{Problem formulation}
\label{sec:problem_formulation}
\subsection{Configuration and flow equations}
\begin{figure}
\begin{center}
\includegraphics[scale = 1]{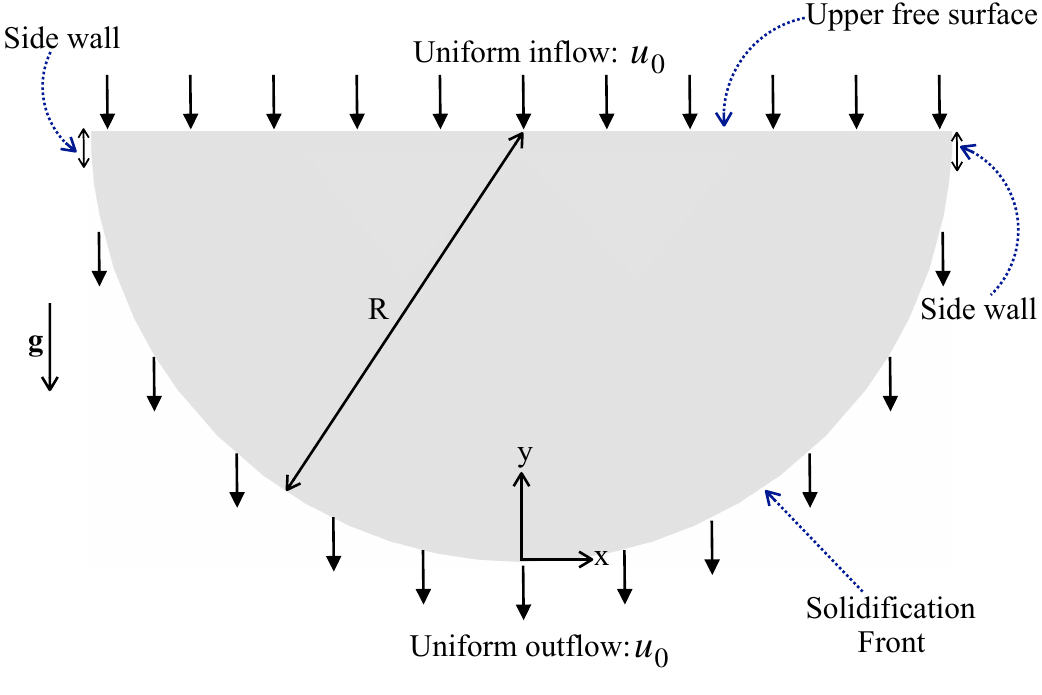}
\end{center}
\caption{The semi-cylindrical geometry with the upper free surface, the side-walls, and the solidification front. The fluid enters at the top and exits through the solidification front with vertical velocity $u_0$.
\label{fig:flow_geo}}
\end{figure}

The problem is mostly modelled as proposed by \cite{Flood:MST1994}.
We consider a stably stratified flow in a cavity with an upper free 
surface where the hot fluid is fed in, and with a cold, porous lower boundary (representing a 
solidification front), as sketched in figure~\ref{fig:flow_geo}. The 
cavity is made of a semi-circular lower boundary representing the actual front, two solid, adiabatic 
side-walls, and is considered infinitely extended in the third direction ($\mathbf e_z$).
Since we focus on the convective mechanisms, detailed solidification mechanisms are not modelled. As 
such, the fluid in the cavity is assumed to remain in a single liquid phase. It is assumed 
Newtonian, incompressible, of density $\rho_0$ at a reference temperature $T_0$, viscosity 
$\nu$, thermal diffusivity $\alpha$, and thermal expansion coefficient $\beta$. Further considering that 
temperature gradients remain moderate, the fluid's motion is described under the Boussinesq 
approximation~\citep{Chandrasekhar:book} and the dynamical equations take the nondimensional form:
\begin{eqnarray}
\frac{\partial {\bf u}}{\partial t}+ ({\bf u} \cdot \nabla) {\bf u} + \nabla p & = & RaPr T {\bf e}_y+Pr \nabla^2 {\bf u},\label{eq:u}  \\
\frac{\partial T}{\partial t}+ ({\bf u} \cdot \nabla) {T} & = &  \nabla^2 T, \label{eq:T} \\
\nabla \cdot {\bf u} & = & 0\label{eq:div_u} , 
\end{eqnarray}
where ${\bf u}=(u,v,w)$ is the velocity vector, $t$ the time, $p$ the modified pressure {including the  buoyancy term accounting for the reference temperature \citep{Chandrasekhar:book,Tritton:book}}, ${\mathbf g=-g\mathbf 
e_y}$ is the gravitational  acceleration. These equations are obtained by normalising lengths by the 
radius of the semi-cylindrical pool $R$, velocities by $\alpha/R$, time by $R^2/\alpha$, pressure by 
$\rho_0(\alpha/R)^2$, and temperature by $\Delta T$.
Here $\Delta T$ is the temperature difference between the hot upper free surface and the cold 
solidification front at the lower boundary. Note that the term involving the reference temperature is absorbed in the pressure gradient, through the modified definition of pressure. { The Prandtl 
number $Pr=\nu/\alpha$ is fixed to $0.02$, a value typical of liquid metals in continuous casting 
processes.}  The Rayleigh number $Ra$ is defined as
\begin{equation}
Ra = \frac{\beta g \Delta T R^3}{\nu \alpha}. \label{eq:Ra}
\end{equation}
The upper boundary at $y=1$ is modelled as a rigid free surface { (standard free-slip boundary condition)}, where incoming fluid at an imposed 
temperature $\Delta T$ is poured with a homogeneous spatial distribution. This is expressed with 
three boundary conditions:
\begin{eqnarray}
\frac{\partial }{\partial y} \mathbf u\times \mathbf e_y=0, \qquad \mathbf u\cdot \mathbf e_y=RePr, \qquad 
T(y=1)=1,
\label{eq:bc_top}
\end{eqnarray}
where the mass flux Reynolds number $Re$, is based on the dimensional feeding velocity $u_0$. In a continuous 
casting process, this velocity would correspond to the casting speed \citep{Flood:MST1994}:
\begin{equation}
Re = \frac{u_0R}{\nu}.
\end{equation}
Thus $Re=0$ corresponds to zero net mass flux, \emph{i.e.}, no fluid crosses through the 
boundaries of the semi-cylindrical pool, while $Re\neq0$ corresponds to non-zero net mass flux, \emph{i.e.} the fluid enters and leaves the domain uniformly through the upper and lower surfaces as in ~\cite{Flood:MST1994}.
The incoming mass flux is exactly cancelled by the flux of fluid being solidified and pulled at the solid 
lower boundary $S$, where solidification imposes the reference temperature:
\begin{eqnarray}
\mathbf u_S\times \mathbf e_y=0 \qquad \mathbf u_S\cdot \mathbf e_y=RePr, \qquad T(y=0)=0.
\label{eq:bc_bot}
\end{eqnarray}
{The kinematic condition expresses that the fluid flows vertically downwards through the 
otherwise no-slip but porous lower boundary. Note that the slight differences between the definitions of the Rayleigh and Reynolds 
numbers given in the introduction and the ones given in this section reflect the difference between 
the geometries discussed there and the specific one we are considering in this paper.}
To ensure consistent boundary conditions for the temperature field at the corners of the domain,
short side-walls of $0.05R$ in height separate the upper and the lower boundaries. 
Impermeable, no-slip boundary conditions for the velocity field and an insulating boundary 
condition for the temperature field are imposed at these { side-walls}. In a real casting 
process, these walls represent the mold. Finally, the infinite extension of the domain in the third 
direction ${\bf e}_z$ {is represented by} periodic boundary conditions 
{for} the velocity and temperature fields. 

\subsection{Linear Stability Analysis \label{sec:lsa_eq}}
The system admits steady base solutions that are invariant along ${\bf e}_z$, similar to 
those found by \cite{Flood:MST1994} (see the detailed topology of these solutions in  \S~\ref{sec:2dsol}).
These may however be unstable to three-dimensional perturbations. Hence, we shall 
detect the corresponding bifurcation by analysing the stability of the base two-dimensional flow to 
infinitesimal three-dimensional perturbations. As such, velocity, pressure, and temperature fields are 
decomposed into the two-dimensional base flow and an infinitesimal three-dimensional perturbation, as
\begin{eqnarray}
{\bf u}(x,y,z,t) & = & {\bf U}(x,y)+{\bf u}^\prime(x,y,z,t), \label{eq:u_decompose} \\
T(x,y,z,t) & = & \bar{T}(x,y)+T^\prime(x,y,z,t), \label{eq:T_decompose} \\
p(x,y,z,t) & = & P(x,y)+p^\prime(x,y,z,t). \label{eq:p_decompose} 
\end{eqnarray}
Substituting equations~(\ref{eq:u_decompose})-(\ref{eq:p_decompose}) into equations~(\ref{eq:u})-(\ref{eq:div_u}) and retaining first-order terms only yields the linearised  equations governing the evolution of  infinitesimal perturbations:
\begin{eqnarray}
\frac{\partial {\bf u}^\prime}{\partial t}+ ({\bf u}^\prime \cdot\nabla) {\bf U}+ ({\bf U} \cdot\nabla) {\bf u}^\prime+\nabla p^\prime & = &RaPrT^\prime {\bf e}_y+Pr\nabla^2{\bf u}^\prime,\label{eq:linear_u}  \\
\frac{\partial T^\prime}{\partial t} + ({\bf u}^\prime \cdot\nabla) \bar{T}+ {\bf U} \cdot\nabla T^\prime & = & \nabla^2 T^\prime, \label{eq:linear_T} \\
\nabla \cdot {\bf u}^\prime & = & 0\label{eq:linear_div_u}.
\end{eqnarray}
The base flow $(\mathbf U, \bar T, P)$ satisfies the same boundary conditions as the main variables 
$(\mathbf u, T, p)$, so that the perturbation {variables} satisfy the homogeneous 
counterpart of the boundary conditions associated { with} the base flow.   
Since the base flow is invariant along ${\bf e}_z$, the general perturbation may be decomposed into normal 
Fourier modes as  
\begin{equation}
{\bf q}^\prime(x,y,z,t) = \sum_{k=-\infty}^{\infty} \hat{{\bf q}}(x,y,t)e^{ikz},
\end{equation}
where ${\bf q}^\prime=({\bf u}^\prime,T^\prime,p^\prime)$ contains all perturbation fields and $k$ is 
the wavenumber along the homogeneous direction ${\bf e}_z$. Further, the absence of the third component of the velocity field in the base flow allows a single phase of the complex Fourier mode to be considered, following~\cite{Barkley:JFM1996},  and others~\citep{Barkley:JFM2002,Sapardi:JFM2017}. The two-dimensionality of the 
base allows us to reduce the three-dimensional perturbation field to a family { of }two-dimensional fields 
parametrised by wavenumber $k$, and computed on the same two-dimensional domain as the base flow.

The linear stability analysis equations shall be solved by means of a time-stepper method: 
defining a linear time evolution operator $\mathcal{A}(\tau)$ for the time integration of 
equations~(\ref{eq:linear_u})-(\ref{eq:linear_div_u}) over time interval $\tau$ as 
\begin{equation}
\hat{{\bf q}}(t+\tau)=\mathcal{A}(\tau)\hat{{\bf q}}(t),
\end{equation}
we solve the eigenvalue problem for operator $\mathcal{A}(\tau)$:
\begin{equation}
\mathcal{A}(\tau)\hat{{\bf q}}_j = \mu_j\hat{{\bf q}}_j.
\end{equation}
Here $\hat{{\bf q}}_j$ denotes the eigenvector of $\mathcal{A}(\tau)$ corresponding to the complex eigenvalue $\mu_j$. The growthrate $\sigma$ and frequency $\omega$ of an eigenmode are related to $\mu$ through 
\begin{equation}
\mu=\exp[(\sigma+i\omega)\tau],\label{eq:mu}
\end{equation}
where the subscript is ignored for brevity. Further, equation~(\ref{eq:mu}) yields,
\begin{equation}
\sigma=\frac{\ln|\mu|}{\tau}; \,\,\,\,\,\,\,\,\,\,\,\,\,\,\,\,\,\,\,\, \omega=\frac{\theta}{\tau},
\end{equation}
with $\mu=|\mu|e^{i\theta}$. An instability occurs if $\sigma>0$, \emph{i.e.}, $|\mu|>1$, while for 
$|\mu|<1$ the corresponding eigenmode is stable. The growing eigenmode may be either oscillatory ($\omega \neq 0$) or non-oscillatory ($\omega=0$). 
The smallest Rayleigh number for which at least one wavenumber $k$ achieves $|\mu|=1$ is the critical Rayleigh number for the onset of instability at a particular value of $Re$, which we shall denote as $Ra_c$.
\section{Computational methods \label{sec:comp}}
\subsection{Numerical set-up}
\label{ssec:num_set}
We perform three different types of numerical computations.
First, steady two-dimensional solutions obtained using Direct Numerical Simulations of (\ref{eq:u}--\ref{eq:div_u}) with associated 
boundary conditions. Two-dimensionality is enforced by setting $\frac{\partial}{\partial z}=0$ and 
$w=0$. Second, the linear stability analysis of three-dimensional perturbations on the two-dimensional 
base flow is carried out, by solving the eigenvalue problem set out in \S~\ref{sec:lsa_eq}. This 
yields the bifurcation points and the structure of{  the} first unstable mode, in the sense of growing $Ra$. 
Third, three-dimensional DNS  are performed in weakly sub- and supercritical regimes for two purposes: 
1) to assess the relevance of the linear stability results,  and 2) to find the nature of the bifurcation. The latter is obtained by { computing the parameters of the Stuart--Landau model from the 3D DNS data}, as 
proposed by \cite{Sheard:JFM2004}.
\begin{figure}
\begin{center}
\includegraphics[scale = 1,trim={0 0 0 1.2cm},clip]{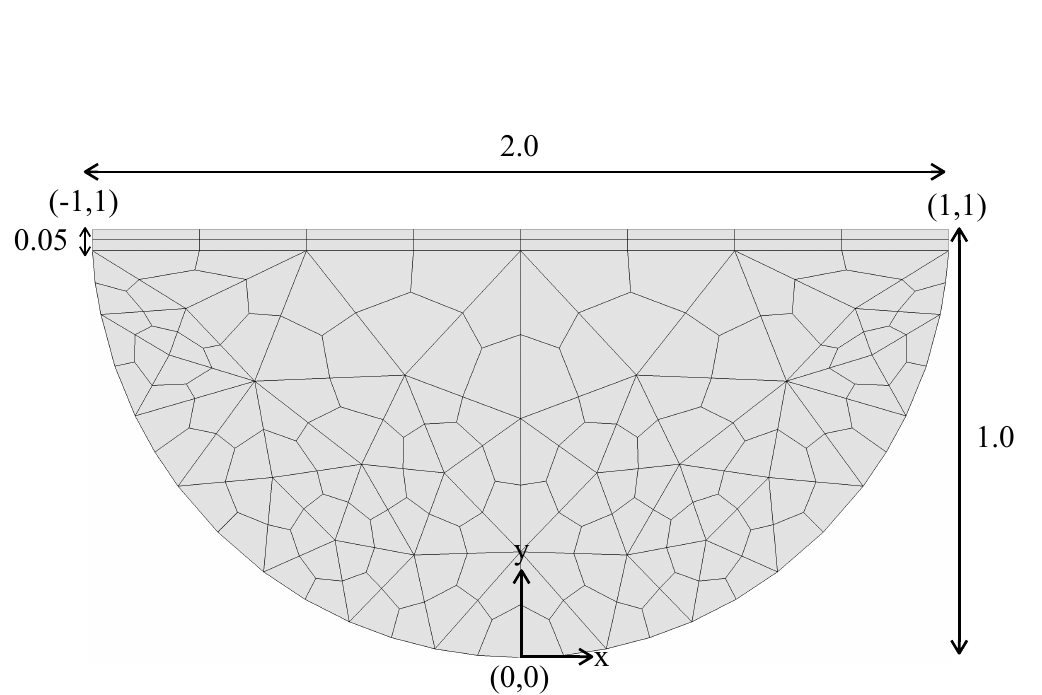}
\end{center}
\caption{Details of the mesh with polynomial order $N=1$. The mesh contains 184 quadrilateral elements.} 
\label{fig:mesh}
\end{figure}

Both the two-dimensional base flow and the evolution of the three-dimensional flow near criticality are obtained by solving  equations~(\ref{eq:u})-(\ref{eq:div_u}) using spectral-element code 
Nektar++~\citep{Cantwell:CPC2015}. In the spectral-element approach, the computational domain is 
partitioned into a mesh of many small subdomains called elements and the variables are projected 
within a polynomial basis within each element, as in the finite element method. The specificity of 
the spectral element method is that ``mesh" refinement is mainly achieved by increasing the order of 
the polynomial basis ($p-$refinement), and that { polynomials} are represented at Gauss--Lobatto points which 
ensure spectral convergence under $p-$refinement. Both two- and three-dimensional DNS are performed on the same spectral element mesh in the $x-y$ plane. 
For the three-dimensional simulations, discretisation in the ${\bf e}_z$ direction relies on a Fourier-based 
spectral method.  { The computational domain extends by $2\pi$ along ${\bf e}_z$, or equivalently, the lowest Fourier mode in the spectral discretisation is always $k=1$.} Figure~\ref{fig:mesh} shows the detail of the two-dimensional $x-y$ mesh with 
polynomial order $N = 1$ generated using the GMSH package~\citep{GeuzaineIJNME:2009}. The mesh is 
composed of 184 quadrilateral elements and  is structured in the rectangular part of the domain close to the upper free surface up to thickness $0.05$, and unstructured in the remaining part of the domain. 
A vertical line along the $y$-axis is imposed to ensure the symmetry of the mesh with respect to that line. At the boundaries, elements are more densely packed than in the bulk, with the ratio between the largest to smallest element's edge size of $4$. Time-stepping relies on a third-order implicit-explicit (IMEX) method~\citep{Vos:IJCFD2011}.

The linear stability analysis is conducted with Open-Source eigenvalue solver DOG (Direct Optimal Growth, \cite{Blackburn:JFM2004,Pitz:JFM2017}), based on a time-stepper method with 
spectral-element discretisation The linearised equations~(\ref{eq:linear_u})-
(\ref{eq:linear_div_u}) are integrated in time using a third-order backward differentiation 
scheme~\citep{Karniadakis:JCP1991}, with the two-dimensional base flow obtained from the DNS. The 
leading eigenvalues and eigenmodes are obtained using the iterative process from the method 
prescribed by \citet{Tuckerman:Springer2000} and \citet{Barkley:IJNMF2008}.

For the two-dimensional direct numerical simulations, the initial condition is set to ${\bf u}=0$, $p=0$, and $T=y$. The three-dimensional direct 
numerical simulations are initiated with the solution obtained for the two-dimensional base flow 
replicated along $\mathbf e_z$, with added white noise of standard deviation of $0.001$ {(as well as $0.01$ and $0.1$ for $Ra<Ra_c$ when the bifurcation is subcritical, \emph{i.e.} $Re=100$ and $Re=110$).}

Cubic spline interpolation of the SCIPY package~\citep{scipy} is used to estimate the Rayleigh number that first produces $\sigma(k)=0$ at different values of $k$. This gives the curve for Rayleigh number vs. $k$, whose minimum yields the estimate of Rayleigh the critical $Ra_c$ and the critical wavenumber $k_c$. Again by cubic spline interpolation, we estimate the value of critical frequency $\omega_c$ for the corresponding $k_c$. The resulting uncertainty on $Ra_c$, $k_c$, and $\omega_c$ remains within $0.01\%$.

\subsection{Code validation}
We first validated the DNS code and the linear stability solver DOG for two-dimensional Rayleigh--B\'enard convection { in a box.  At the top and bottom plates, a no-slip  boundary condition  for the velocity field and a conducting boundary condition for the temperature field are imposed. Periodic 
boundary conditions are applied at the side-walls for both the fields.}  DNS were validated by calculating 
the Nusselt number 
\begin{equation}
Nu = \frac{\alpha\frac{\Delta T}{R}+ \langle wT\rangle }{\alpha\frac{\Delta T}{R}},
\end{equation}
where $ \langle  \rangle$ represents the volume average.
$Nu$ measures the ratio of the total (convective plus conductive) heat flux to the conductive heat flux. It was computed for $Pr=0.71$ and $Ra=5000$ in a two-dimensional box and the relative error between the Nusselt number computed from our DNS and that of~\cite{Clever:JFM1974} was $0.09$\%. For the linear stability analysis, we recovered  the known values of Rayleigh critical ($Ra_c=1707.7$) and critical wavenumber ($k_c=3.116$) found for instance in~\cite{Chandrasekhar:book}, to within an uncertainty of $0.002$\%.
For both DNS and linear stability analysis, a rectangular mesh of $100$ structured quadrilateral elements with polynomial order $N=9$ was used. 

\subsection{Convergence tests}
\label{ssec:convergence_test}
Since the spectral element discretisation is only used in the $x-y$ plane, we performed a convergence 
test on the base two-dimensional flows and the eigenvalue problem based on the two-dimensional domain. 
This test was performed for each value of $Re$ we considered throughout this work, both on the 
DNS and on the leading eigenvalue returned by the linear stability analysis at $k$ for which $\sigma(k)$ is maximum. 
Table~\ref{tab:convergence_Re_500} shows an example for the accuracy of the eigenvalue calculations as a 
function of the polynomial degree $N$ for $Re=500$, the highest value of $Re$ considered in 
this work. The leading eigenvalue for $Ra=2\times 10^6$ and $k=13$ is real and linearly unstable. 
We increased $N$ until the eigenvalue converged to a precision of $5$ significant figures, which is $N=14$  for this case.  This way the same level of precision was achieved for all results presented in this work. { The time-step was kept constant for all three types of numerical calculations, so that 
the maximum local Courant number $C_{\rm max}$ remain below unity, and the Courant-Friedrich-Levy 
condition be strictly satisfied everywhere in the domain, at all time. For instance, at $Re=500$, $Ra=2\times 10^6$, and $N=16$, the time-step is $10^{-6}$, which yields a maximum Courant number of $C_{\rm max}=0.05$.}

\setlength{\tabcolsep}{20pt}
\begin{table}
\begin{center}
\def~{\hphantom{0}}
\begin{tabular}{c c c}
$N$  & $|\mu_{max}|$   &   Relative error (\%) \\[2 mm]
$9$   & $1.0361$ & $0.55668$\\
$10$   & $1.0399$ & $0.19196$\\
$11$   & $1.0413$ & $0.05759$\\
$12$   & $1.0415$ & $0.03839$\\
$13$   & $1.0418$ & $0.00960$\\
$14$   & $1.0419$ & ---\\
$15$   & $1.0419$ & ---\\
$16$   & $1.0419$ & ---\\
\end{tabular}
\caption{Dependence of leading eigenvalues on the polynomial order $N$. Leading eigenvalues computed on the mesh at  $Re=500$, $Ra=2 \times10^6$, and $k=13$ are provided.  The relative error is to the case of the highest polynomial order ($N=16$).}
\label{tab:convergence_Re_500}
\end{center}
\end{table}

Furthermore, some supercritical cases (for example $Re=200$; $Ra=10^6$) are unstable to 
time-periodic, two-dimensional perturbations. In these cases, a DNS of the base flow over the entire 
two-dimensional domain does not converge to the base steady solution. Since the perturbation breaks a 
symmetry with respect 
to the $x=0$ plane, we found the steady state by performing a DNS on the  $x\leq0$ half of the domain, 
implementing symmetry boundary conditions for the velocity (Neumann boundary condition on 
$u$, Dirichlet boundary condition on $v$) and the temperature (Neumann boundary condition) at 
$x=0$. 
The solution over the full domain was then built by symmetry~\citep{Mao:PF2014}. We checked that the relative error in 
the magnitude of the most dominant mode for $Re=200$, $Ra=10^5$, and $k=7$ obtained from 
LSA on base flows calculated on the full domain and the half-domain was less than 0.01\%. Thus, the use of  
half-domain calculation for the base flow is justified, and also computationally cheaper.

For the three-dimensional DNS, the dependence on the number of Fourier modes $N_f$ was tested by 
computing the total kinetic energy of the 3D flow over the entire domain. The relative error in the 
kinetic energy of the supercritical steady state at $Re=200;Ra=1.1\times 10^5$ between computations at $N_f=32$ 
and $N_f=64$ was $0.01 \%$. Similar results were obtained for other values of  $Re$, and thus 32 
Fourier modes were employed for all cases. Note that the three-dimensional simulations are performed 
only up to $Re=200$, as the simulations for higher values of $Re$ become prohibitively expensive.
A summary of all cases investigated is provided at the end of the paper in table \ref{tab:summary}.

\section{Two-dimensional base flows \label{sec:2dsol}}
\subsection{Dynamics of base flow with zero net mass flux ($Re=0$)}
\begin{figure}
\begin{center}
\includegraphics[scale = 0.8,trim={0 0 0 1.2cm},clip]{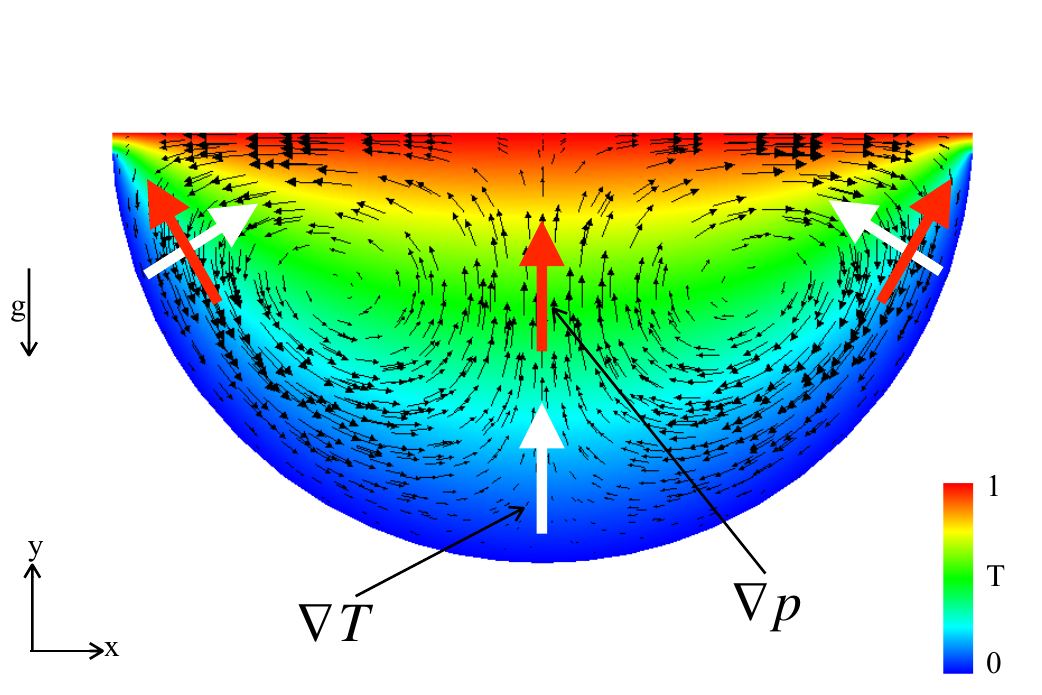}
\end{center}
\caption{Two-dimensional base flow at $Re=0$ and $Ra=1$. The pressure gradient ($\nabla p$) and the temperature gradient ($\nabla T$) are indicated by red and white arrows, respectively. These gradients form an angle at the upper corners which results in a baroclinic imbalance. The density gradient is related to the temperature gradient as: $\nabla \rho=-\beta^{-1}\nabla T$ in dimensional variables. {Note that in this example, the modified pressure is predominantly determined by the buoyancy, as it includes the buoyancy term corresponding to the reference temperature}.} 
\label{fig:baro}
\end{figure}

To perform a linear stability analysis, we first need to obtain the two-dimensional base flow. We 
first focus on the behaviour of two-dimensional base flow with zero net mass flux. Figure~\ref{fig:baro} 
displays the density plots of the temperature  field and the velocity vector field ${\bf u}$ in  
the $x-y$ plane for $Re=0$ and $Ra=1$. Despite a ``stable" stratification (\emph{i.e.} the light 
fluid is mostly on the top of the heavy one), we observe prominent convective rolls.
This motion is driven by the misalignment of the pressure gradient and the temperature gradient, 
which generates a baroclinic imbalance. This effect is best seen from the dimensional vorticity 
equation, where the curl of the pressure term yields a vorticity source term proportional to the 
baroclinic vector $1/\rho(T)^2 \nabla \rho(T) \times \nabla p$~\citep{Vallis:Book}, where the density 
gradient is $ \nabla \rho(T)=-\nabla T/\beta$ in dimensional variables. In configurations where the 
density gradient and the pressure gradient are aligned, as in plane Rayleigh--B\'enard convection, this term 
cancels out. Near the semi-circular 
boundary of the domain, the temperature gradient, hence the density gradient is radial, whereas the 
pressure gradient (dominated by buoyancy) is tilted.  While these directions coincide near the centre of the domain (around $x=0$), the 
angle they form increases to a maximum near the side-walls. At these locations, the baroclinic vector 
acts as a source of $z-$vorticity that drives a strong downward jet along the circular boundary. The 
rolls form as a result of the left and right jets meeting at $x=0$ where the flow returns.
Since only viscous friction opposes this motion, baroclinic imbalance drives a non-zero base flow, even 
at arbitrary low stratifications.

Interestingly, if the curvature of the lower boundary was continuously decreased to 0, the angle between 
the density and pressure gradient would progressively decrease to 0 too. In the limit of a flat 
lower boundary, the Rayleigh-B\'enard configuration {would be recovered in a channel of height determined by the side-walls, even though these would be pushed to infinity. However, the stratification would be stable and since the baroclinic vector would cancel out in this limit, the base flow would be still and stable, with a linear temperature gradient in the bulk.}

\begin{figure}
\begin{center}
\includegraphics[width=\textwidth]{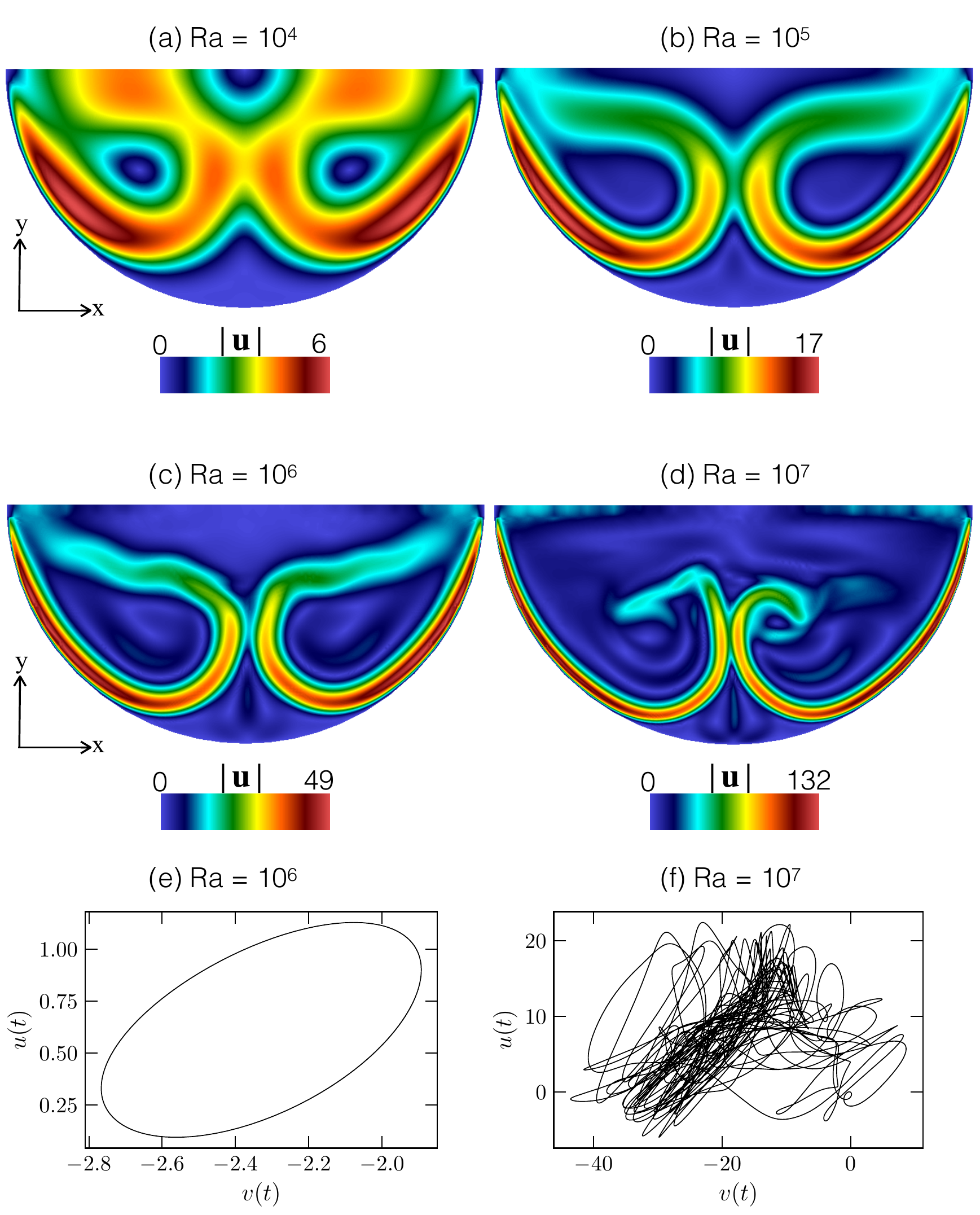}
\end{center}
  \caption{Two-dimensional base flow with zero net mass flux ($Re=0$) at (a) $Ra=10^4$, (b) $Ra=10^5$, (c) $Ra=10^6$, and (d) $Ra=10^7$. Colours represents the magnitude of the velocity field. Figures (e) and (f) display Lissajous graphs in the $(u,v)$ plane for $Ra=10^6$ and $Ra=10^7$ { measured at $(x,y) = (-0.5, 0.6)$}, respectively exhibiting periodic and chaotic states.}
\label{fig:2D_base_Re}
\end{figure}

We now gradually increase the Rayleigh number from $Ra=10^4$ to $Ra=10^7$ with $Re=0$. As expected, 
the velocity field strengthens at the wall with the increase in Rayleigh number, since the intensity of 
the density gradient in the baroclinic vector increases (see figure~\ref{fig:2D_base_Re}(a)-(d)).  
While the two-dimensional solution remains steady for $Ra=10^4$ and $Ra=10^5$, it becomes 
time-periodic for $Ra=10^6$, and chaotic for $Ra=10^7$. The periodic and chaotic nature of the 
solutions for $Ra=10^6$ and $Ra=10^7$ are illustrated by the Lissajous graphs of the 
two components of velocities in figures~\ref{fig:2D_base_Re}(e) 
and ~\ref{fig:2D_base_Re}(f). 

\setlength{\tabcolsep}{10pt}

\subsection{Base solution with non-zero mass flux  ($Re>0$) }
\label{sec:2D_base_non_zero_mass_flux}
\begin{figure}
\begin{center}
\includegraphics[width=\textwidth]{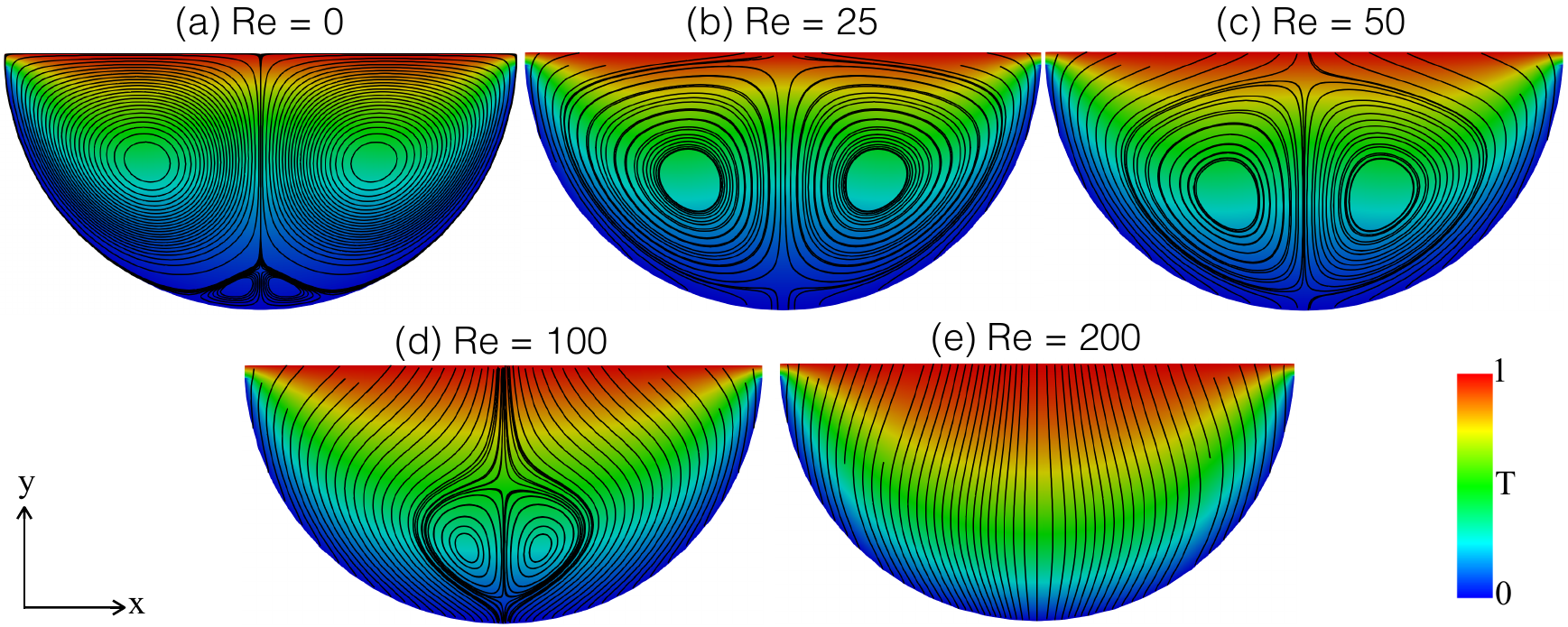}
\end{center}
\caption{Streamlines of the steady two-dimensional base flow and temperature field for $Ra=10^4$ at (a) $Re=0$, (b) $Re=25$, (c) $Re=50$, (d) $Re=100$, and (e) $Re=200$.} 
\label{fig:streamlines}
\end{figure}

One of the key ingredients in the problem we consider is the through-flow. To illustrate its effect on 
the base flow, we computed the two-dimensional base flows for a range of Reynolds number from 0 to 200, 
at a fixed value of $Ra$ of $10^4$ (for the purpose of seeking the critical Rayleigh number for the 
linear stability, the base flow will have to be recalculated  every time either $Ra$ or $Re$ is changed).
 Figure~\ref{fig:streamlines} shows the streamlines of the velocity field and the density plots of the 
temperature field for the two-dimensional base flow. At $Re=0$ the flow consists of the pair of primary 
vortices and the pair of secondary vortices which are located at the  bottom of the cavity, as discussed in the previous section. At low values of $Re$ ($Re=25$), the primary vortices are slightly displaced 
downwards and secondary vortices are suppressed under the combined action of the through-flow and the 
confinement at the lower boundary. The size of the primary vortices increases as a result, 
and the convection associated to them is slightly enhanced. However, with a further increase in $Re$, 
the same mechanism incurs a suppression of the primary vortices from the top and a reduction of their 
size. At $Re=200$ the primary vortices have completely disappeared. Thus, the through-flow enhances 
convection at low Reynolds numbers (below 25, in the set of Reynolds numbers we explored), but 
suppresses it at higher Reynolds numbers. 

{
\subsection{Vertical heat flux}
\label{ssec:heat_flux}
The efficiency of the convection is estimated by quantifying the vertical heat flux. For this purpose, we compute the Nusselt number at the inlet of the cavity defined as 
\begin{equation}
Nu =  \frac{- \left\langle \left(\frac{\partial T}{\partial y}\right)_{y=1} \right\rangle_{\rm Convective} +  (vT)_{y=1}}{- \left\langle \left(\frac{\partial T}{\partial y}\right)_{y=1} \right\rangle_{\rm Conductive} +  (vT)_{y=1}}, \label{eq:Nu_Inlet}
\end{equation}
and analyse its  variations with the Rayleigh number for several fixed values of $Re$. Here the reference conductive states, is chosen as a uniform downward flow at non-dimensional velocity $RePr$, without rolls. This way, the Nusselt number in (\ref{eq:Nu_Inlet}) measures the enhancement of heat transfer due to 
the convective flow inside the cavity, and not that due to the average downward fluid motion. The  
temperature distribution for this reference state is obtained by solving the steady advection-diffusion 
equation,
\begin{equation}
\nabla^2T - ({RePr {\bf e}_y \cdot \nabla})T=0,
\end{equation}
with the same boundary conditions for the temperature as for the base flow (\ref{eq:bc_top}) and 
(\ref{eq:bc_bot}). 
The variations of $Nu$ with the Rayleigh number for the two-dimensional base flow are shown in 
figure~\ref{fig:Nu}(a). For $Re=0$, the Nusselt number monotonically increases with $Ra$. By contrast, 
for $Re\neq 0$, $Nu$ first decreases before reaching a minimum, and it goes below unity for $Re=75$ to $Re=200$. The decrease of $Nu$ below unity occurs as the 
two rolls form and gain in intensity. As they do so, they redistribute heat laterally in the upper part 
of the pool. As they grow in size, they do so ever closer to the top boundary and oppose the downward 
temperature gradient near the corners, and hence the heat flux through the upper boundary.  When 
the convective rolls have grown to occupy the entire cavity, however, their shape does not evolve 
anymore as $Ra$ is further increased. In this regime, the heat flux near the corners saturates and the 
main effect of the rolls is to convey cold fluid upwards near $x=0$. The cooling of the region near 
upper boundary that ensues leads $Nu$ to increase again and soon exceed unity.  There are relatively few instances where convection reduces, rather than enhances heat transfer (\emph{i.e.} $Nu<1$). It has, for example, been observed in the Rayleigh--B\'enard convection of homeotropically aligned nematic liquid crystal 
\citep{Thomas:PRE1998}.

\setlength{\tabcolsep}{10pt}
\begin{figure}
\centering
   \begin{tabular}{ll}
    (a) & (b)\\
  \parbox{0.45\textwidth}{\includegraphics[width=0.49\textwidth]{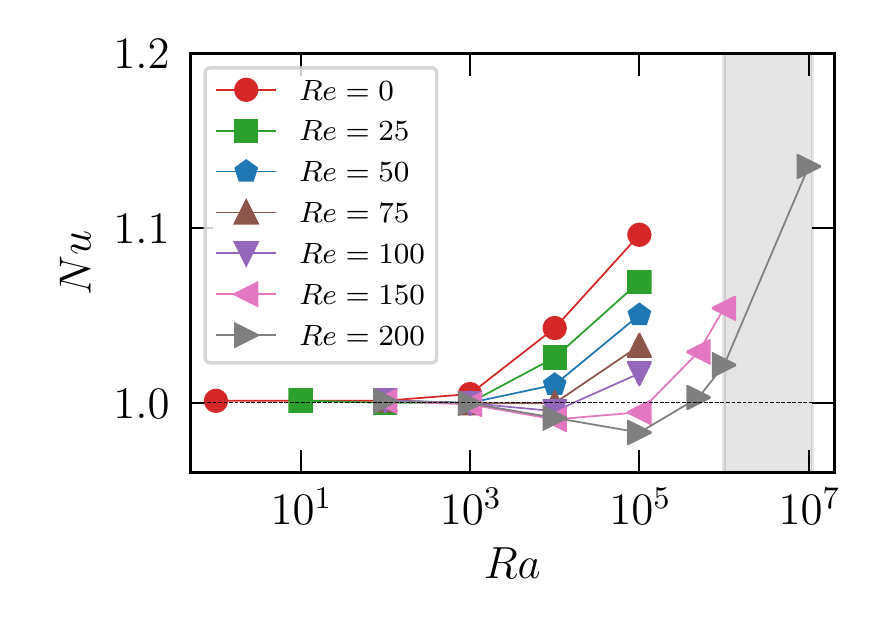}} & 
    \parbox{0.45\textwidth}{\includegraphics[width=0.45\textwidth]{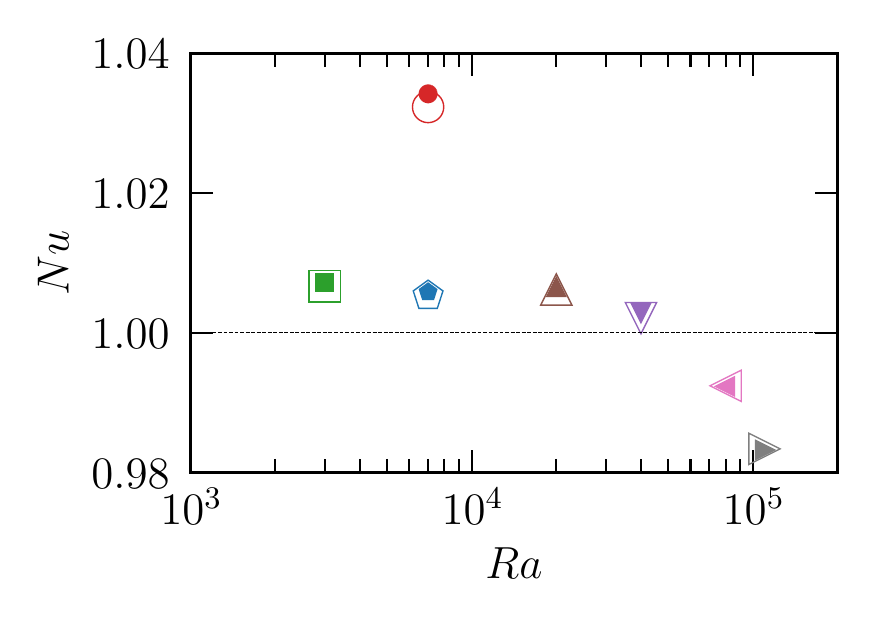}}\\ 
   \end{tabular}
  \caption{Nusselt number $Nu$ calculated at the inlet: (a) For two-dimensional base flow for various values of $Ra$. The shaded region represents the unsteady case; (b) A comparison between $Nu$ for two-dimensional (represented by solid symbols) and three-dimensional (represented by hollow symbols) DNS at the same value of the Rayleigh number. Points are obtained near criticality (hence, for different values of $Re$).}
\label{fig:Nu}
\end{figure}
}

\section{Stability analysis \label{sec:lsa}}
\subsection{Growth rates and Eigenvalue spectra}
\label{ssec:growth_rate}
\begin{figure}
\centering
    \begin{tabular}{ll}
    (a) $Re=0$ & (d) \\
    \parbox{0.5\textwidth}{\includegraphics[width=0.55\textwidth]{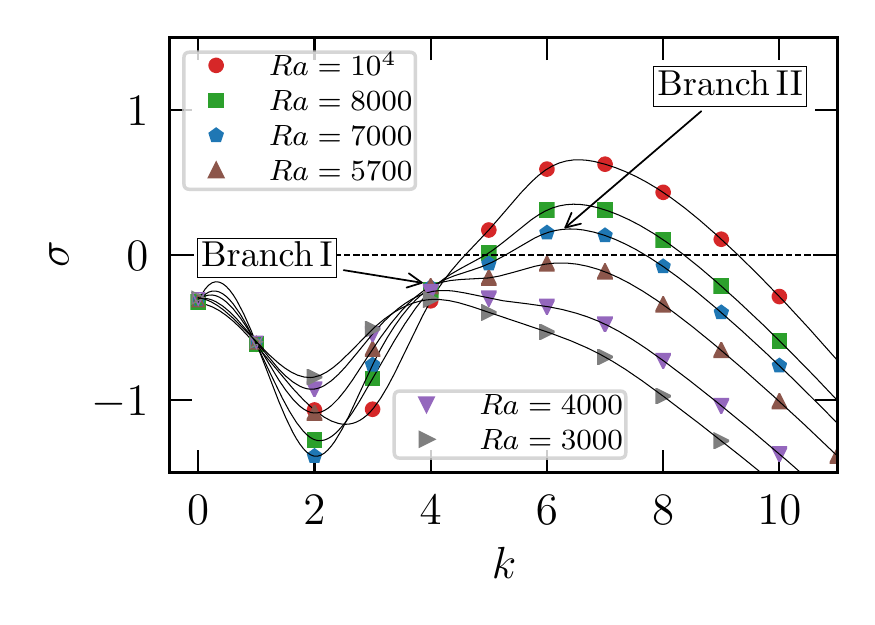}} &
    \parbox{0.39\textwidth}{\includegraphics[width=0.4\textwidth]{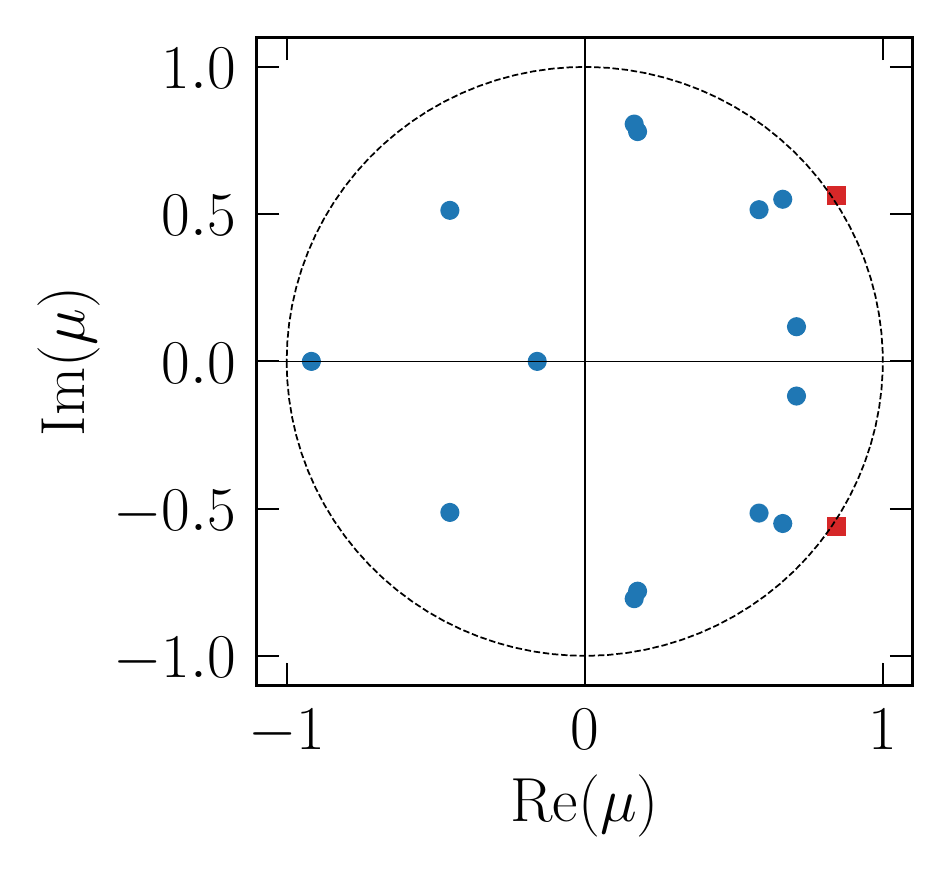}} \\
    (b) $Re=100$ & (e) \\
    \parbox{0.5\textwidth}{\includegraphics[width=0.55\textwidth]{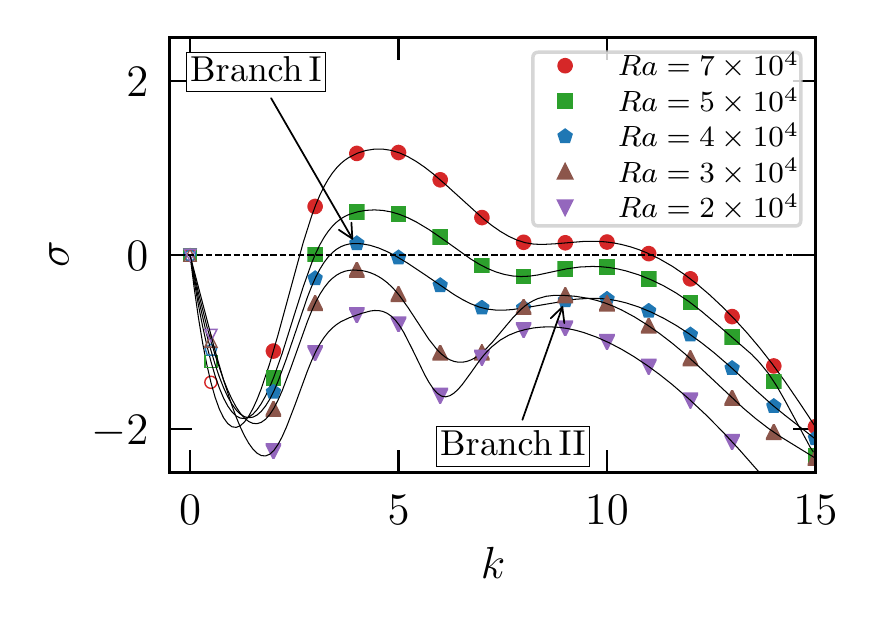}} &
    \parbox{0.39\textwidth}{\includegraphics[width=0.4\textwidth]{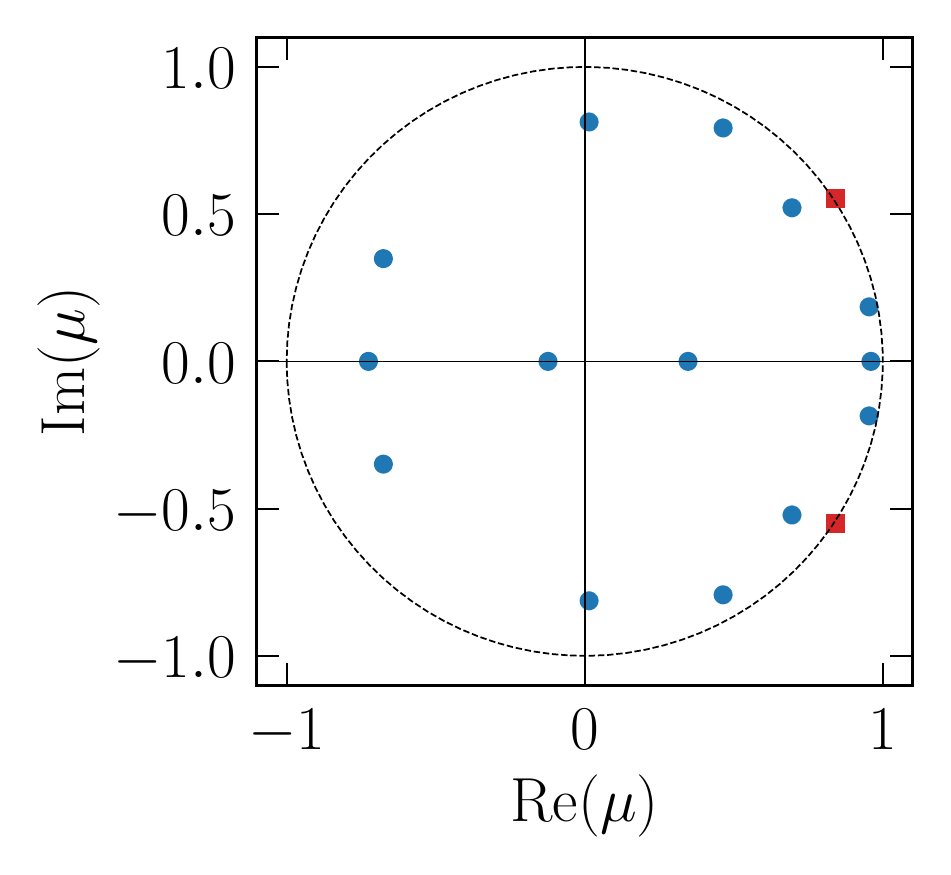}} \\
    (c) $Re=200$ & (f) \\
    \parbox{0.5\textwidth}{\includegraphics[width=0.55\textwidth]{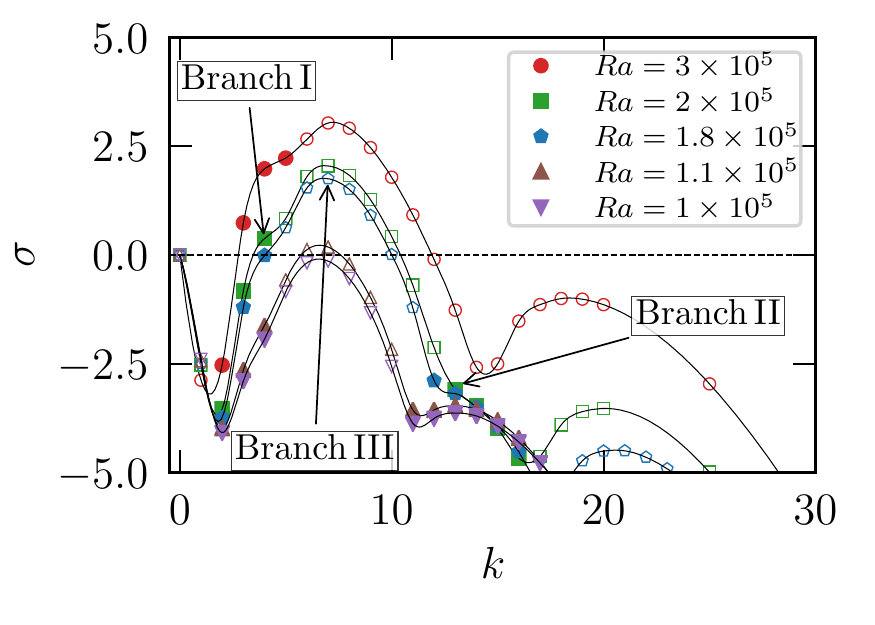}} &
    \parbox{0.39\textwidth}{\includegraphics[width=0.4\textwidth]{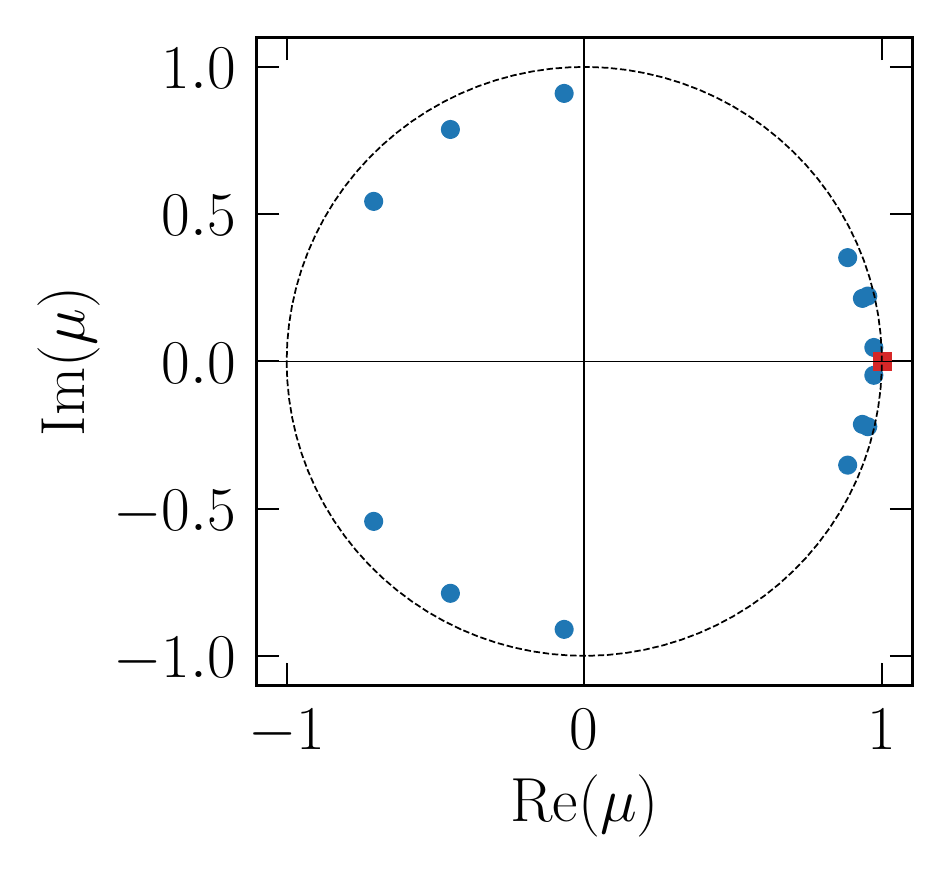}} \\
  \end{tabular}
  \caption{Left column: Growthrates of leading eigenmodes as functions of the wavenumber $k$ for (a) $Re=0$ and $Ra\leq10^4$, (b) $Re=100$ and $Ra\leq 7 \times 10^4$ and (c)  $Re=200$ and $Ra \leq 3 \times 10^5$. Solid symbols represent real leading eigenvalues, while hollow symbols represent complex-conjugate pairs of non-real leading eigenvalues. Right column: Eigenvalue spectra for marginally supercritical cases 
for (d) $Re=0$, $Ra=7 \times 10^3$ and $k=6$,  (e)  $Re=100$, $Ra=4 \times 10^4$ and $k=4$ and (f) $Re=200$, $Ra=1.1 \times 10^5$ and $k=7$. \textcolor{C0}{$\bullet$} and \textcolor{C3}{$\filledmedsquare$} represent stable eigenvalues and the first unstable eigenvalue, respectively.}
\label{fig:growth_rate}
\end{figure}

We now turn to the linear stability of the steady two-dimensional states found in \S~\ref{sec:2dsol}. We start by analysing the dependence of the perturbation growth $\sigma$ on the Rayleigh number $Ra$, 
wavenumber $k$, and mass flux Reynolds number $Re$. The left column of figure~\ref{fig:growth_rate} shows
 the growthrate $\sigma$ as a function of wavenumber $k$ for $Re=0$, $100$, and $200$. 
For a particular value of $Re$, the growthrate $\sigma(k)$ is computed for a range of Rayleigh numbers until supercritical regimes are encountered. The right column of figure~\ref{fig:growth_rate} shows the eigenvalue spectra for $Re$ (corresponding to the left column) near to the onset of instability, in marginally supercritical regime. 
\setlength{\tabcolsep}{10pt}
\setlength{\tabcolsep}{10pt}
\begin{figure}
\centering
   \begin{tabular}{ll}
    (a) & (b)\\
  \parbox{0.45\textwidth}{\includegraphics[width=0.49\textwidth]{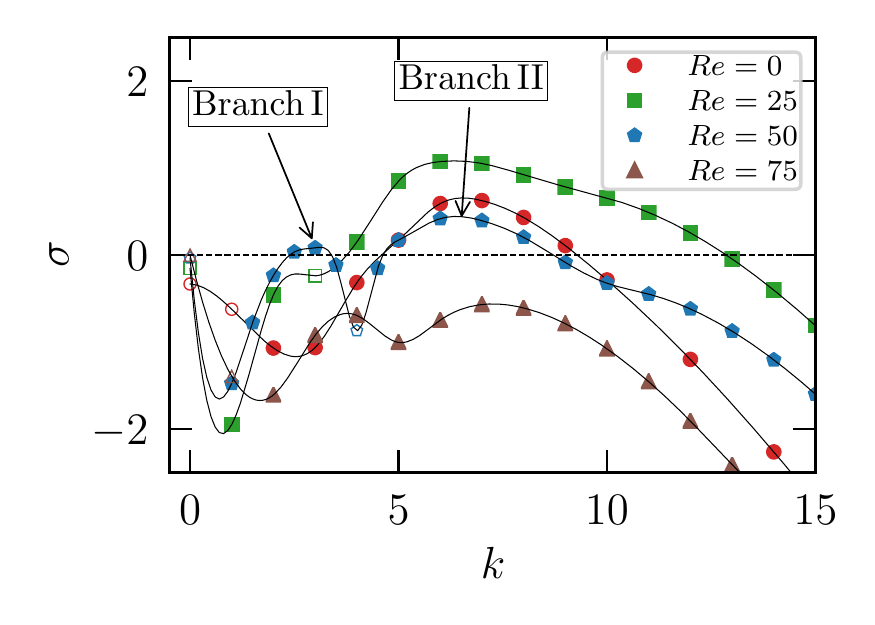}} & 
    \parbox{0.45\textwidth}{\includegraphics[width=0.45\textwidth]{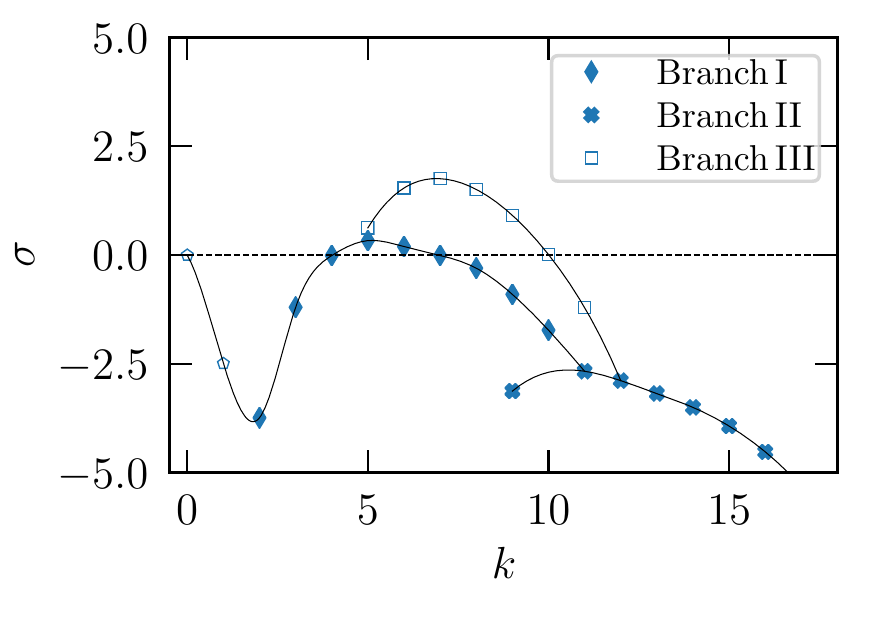}}\\ 
   \end{tabular}
  \caption{(a) Growthrates of the leading eigenmodes at $Ra=10^4$ for $Re=0$, $25$, $50$, and $75$.  (b) Growthrates of the eigenmodes associated to branches I, II and III for $Re=200$ and $Ra=1.8 \times 10^5$. Each curve $\sigma(k)$ in figure~\ref{fig:growth_rate} corresponds to the absolute maximum growthrate over all three branches for each $k$ .}
\label{fig:Dual_branch}
\end{figure}
For $Re=0$, all eigenmodes are oscillatory ($\omega\neq0$) over the entire range of values of $Ra$ we 
investigated. 
At low wavenumber $k\simeq4$, a local maximum in $\sigma(k)$ is observed. We denote the set of 
eigenvectors forming this maximum as ``Branch I". These remain stable for all values of $Ra$ we 
investigated. A second, absolute maximum exists around $k\simeq6$ and we shall label the corresponding set of modes as ``Branch II". The mode associated to this maximum becomes unstable for 
$Ra=7\times 10^3$.
The corresponding pair of complex conjugate eigenvalues are shown in figure~\ref{fig:growth_rate}(d) as 
they cross the unit circle that separates stable from unstable eigenmodes.
Branches I and II (as well as Branch III discussed further up in this section, and which is dominant in this case) are illustrated more 
extensively on the example of $Re=200$, $Ra=1.8 \times 10^5$ in figure \ref{fig:Dual_branch}(b).

\begin{figure}
\centering
   \begin{tabular}{ll}
    (a) & (b)\\
  \parbox{0.45\textwidth}{\includegraphics[width=0.49\textwidth]{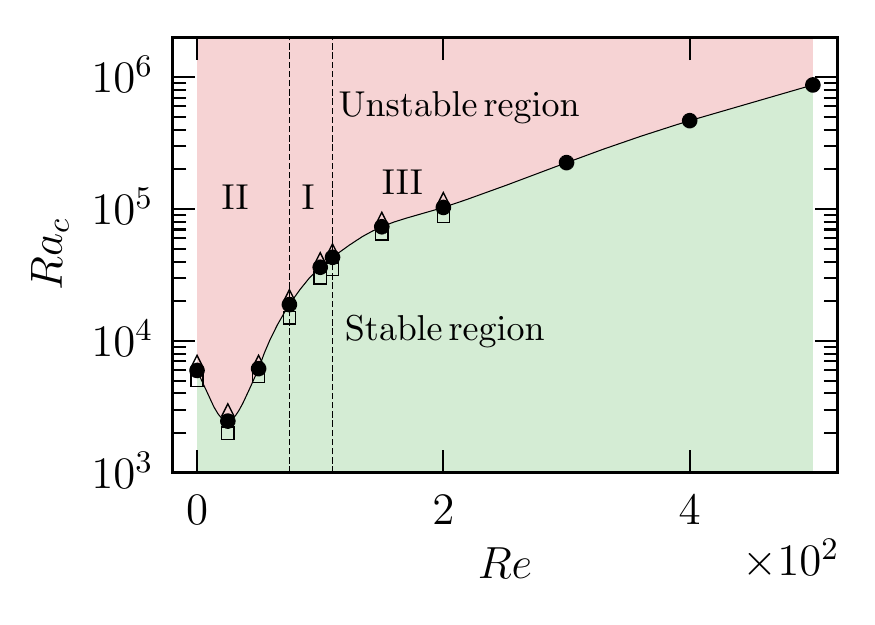}} & 
    \parbox{0.45\textwidth}{\includegraphics[width=0.49\textwidth]{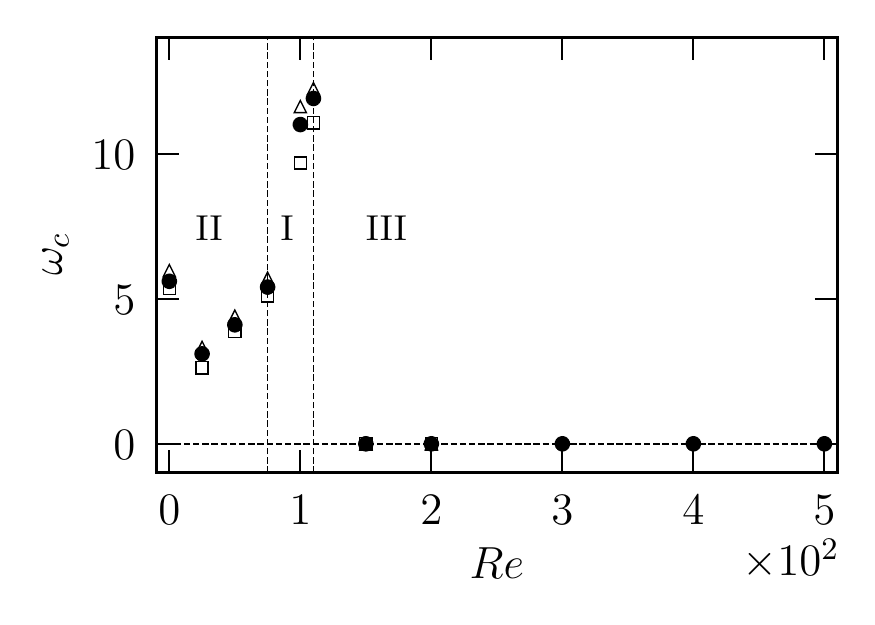}}\\ 
       \end{tabular}
  \begin{tabular}{l}
    (c) \\  
    \parbox{0.45\textwidth}{\includegraphics[width=0.49\textwidth]{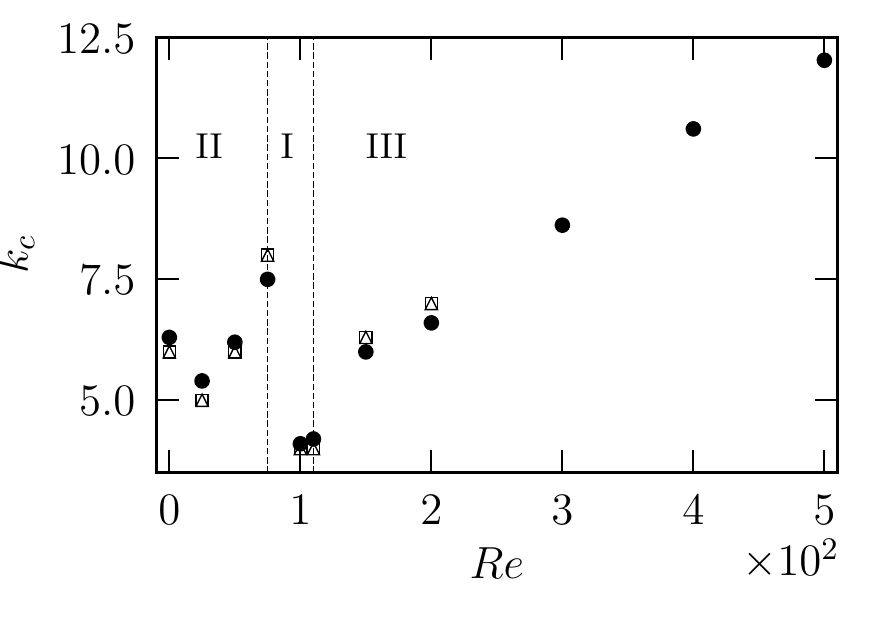}} \\
   \end{tabular}
  \caption{(a) Critical Rayleigh number, (b) critical frequency, and (c) critical wavenumber as a 
function of $Re$. In (a) green (red) regions below (above) the curve represents flow regimes that are linearly stable 
(unstable) to two or three-dimensional perturbations. `\textcolor{black}{$\bullet$}'  data obtained from the linear stability analysis. Three-dimensional direct numerical simulations up to $Re=200$ at slightly subcritical (\textcolor{black}{$\square$}) and slightly supercritical (\textcolor{black}{$\bigtriangleup$) values of $Ra$}.}
\label{fig:Dependence_on_Re}
\end{figure}

When $Re$ is increased from 0, the growthrate first increases for all $k$ (keeping the 
same value of $Ra$). Consequently, the critical Rayleigh number $Ra_c(Re)$ initially decreases 
up to $Re=25$  (see figure~\ref{fig:Dependence_on_Re}(a)). The physical reason can be traced 
to the structure of the two-dimensional base flow: increasing $Re$ from 0 suppresses the secondary 
vortices near $x=0$. These give way to the main bulk cells which grow in size (see figure~\ref{fig:streamlines}(b)). This implies that 
the effective flow lengthscale increases, and so does the effective Rayleigh number. Consequently, a 
lower Rayleigh number is sufficient to trigger the instability.

For $Re>25$, the effect is reversed. $\sigma(k)$ decreases for all $k$ and the corresponding 
critical Rayleigh number $Ra_c(Re)$ increases. This time, the effect originates in the reduction in 
the size and intensity of the main cells in the base flow, due to their suppression by the inflow near the top 
boundary at $y=1$ (see figure~\ref{fig:streamlines}(c)). The decrease of $\sigma(k)$ is however not uniform and the maximum of $\sigma(k)$ associated to branch I increases in value compared to that associated to branch II. This effect is best illustrated in figure~\ref{fig:Dual_branch}(a), which displays 
$\sigma(k)$ vs. $k$ at fixed value of Rayleigh number ($Ra=10^4$) for $Re=0$, $25$, $50$ and $75$.
From $Re=100$  onwards, the most unstable mode becomes associated to branch I. From this point on, $Ra_c$ continues to increase with $Re$ but more slowly than for $Re\leq100$.

Up to $Re=110$, all calculated unstable eigenmodes are oscillatory. From $Re=150$,
a third branch (III) appears very close to branch I, with the specificity that all associated 
modes are non-oscillatory (\emph{i.e.} $\omega=0$). For $Re\geq150$, the mode associated to the 
maximum growthrate in branch III becomes dominant and the onset of the instability occurs through a 
non-oscillatory mode. 
The real eigenvalue associated to the single fastest growing mode is represented in the eigenvalue spectra for $Re=200$, $Ra=1.1 \times 10^5$ at $k=7$ in 
figure~\ref{fig:growth_rate}(f). At this point, the main cells in the base flow have disappeared. 
Hence, the transition from the oscillatory to the non-oscillatory instability is associated to a 
fundamental change in the nature of the base flow, as it switches from a recirculation-dominated 
topology to one dominated by the through-flow. As such, this transition separates a 
{buoyancy-driven} regime and a hydrodynamic one.

\setlength{\tabcolsep}{10pt}
\subsection{Variations of $\omega_c$ and $k_c$ with $Re$}
\label{ssec:Dependence_on_Re}
The variations with $Re$ of the critical frequency $\omega_c$ and wavelength $k_c$ at the onset of 
instability reflect the transition between branches I, II and III, whose co-existence  is illustrated 
in figure~\ref{fig:Dual_branch}(b). 
As $Re$ increases, the 
variations of the frequency and wavelength associated to the maximum of each of the branches differ: 
for $Re\leq100$, branch II dominates and  over this interval, both $\omega_c$ and $k_c$ follow the 
non-monotonous variations of $Ra_c$ observed in the previous section. 
The switchover from branch II to branch I observed at $Re=100$ translates into a discontinuity 
in the variations of both $\omega_c$ and $k_c$. While $\omega_c$ practically doubles between $Re=75$ and
$Re=100$, $k_c$ drops by half over the same interval (evaluating the exact amplitudes
of these discontinuities would require a prohibitively large number of simulations). However, both 
$\omega_c$ and $k_c$ subsequently increase over the interval of dominance of branch I. 
The next discontinuity occurs at $Re=150$, at which point branch III becomes dominant at the expense 
of branch I. Since branch III modes are non-oscillatory, $\omega_c$ drops to zero. At the same time, the 
 $k_c$ jumps up to higher values and continues to increase with $Re$ beyond $Re=150$.
 
Discontinuities in length-scale at the onset of instabilities are frequently observed when convection is 
combined with forces other than buoyancy and viscosity. {Usually, the discontinuity appears when a parameter representing the ratio of two 
of the forces in the system is varied, and it reflects the transition between 
different unstable modes. }
In rotating magnetohydrodynamic (MHD) convection, {for example},
a transition occurs between the thin convective plumes favoured by fast rotation and the large convective rolls, favoured by the Lorentz force. As in the present case, each of these patterns 
corresponds to a distinct branch of eigenmodes of the stability problem. The 
transition between them
takes place at a critical value of the ratio between these forces. 
The corresponding change in length-scale has been experimentally observed to reach an order of magnitude 
\citep{nakagawa1957_prsa}, in agreement with theoretical predictions \citep{Chandrasekhar:book,aujogue2015_pf}. Similar phenomena are also observed in mixed convection in magnetic fields, at the changeover 
between convection dominated regimes and shear-dominated ones \citep{vo2017_prf}. While in rotating 
magnetoconvection, the transition only involves non-oscillatory modes, it only involves oscillatory ones 
in mixed MHD convection. The transition resembling most the one observed here, between oscillatory and 
non-oscillatory modes, and with a discontinuity in wavelength, was observed when decreasing the Prandtl number in rotating convection \citep{clune1993_pre}.

Finally, in the range of larger Reynolds numbers, an asymptotic behaviour associated to branch III 
emerges, where $k_c$ scales linearly with $Re$ as $k_c=(0.018 \pm 0.001) Re + (3.2 \pm 0.2)$.

\setlength{\tabcolsep}{20pt}
\begin{table}
\begin{center}
\def~{\hphantom{0}}
\begin{tabular}{c c c c}
$Re$  & $Ra_c$   &   $k_c$ & $\omega_c$ \\[2 mm]
$0$   & $5.975 \times 10^3$ & $6.3$ & $5.6$\\
$25$   & $2.467 \times 10^3$ & $5.4$ & $3.1$\\
$50$   & $6.168 \times 10^3 $ & $6.2$ & $4.1$\\
$75$   & $1.887 \times 10^4 $ & $7.5$ & $5.4$\\
$100$   & $3.621 \times 10^4$ & $4.1$ & $11.0$\\
$110$   & $4.307 \times 10^4$ & $4.2$ & $11.9$\\
$150$   & $7.345 \times 10^4$ & $6.0$ & $0$\\
$200$   & $1.029 \times 10^5$ & $6.6$ & $0$\\
$300$   & $2.250 \times 10^5$ & $8.6$ & $0$\\
$400$   & $4.680 \times 10^5$ & $10.6$ & $0$\\
$500$   & $8.737 \times 10^5$ & $12.0$ & $0$
\end{tabular}
\caption{Critical Rayleigh number, corresponding wavenumber along the homogeneous direction and frequency at the onset of instability for $Re$ ranging from zero to $500$.}
\label{tab:critical_values}
\end{center}
\end{table}
\subsection{DNS near criticality}
\label{ssec:nonlinear_analysis}
To assess the relevance of the linear stability analysis, we perform DNS of the two-dimensional flow 
perturbed by white noise as described in  \S~\ref{ssec:num_set}, for each value of $Re$ up to 
200, at slightly subcritical ($r_c<0$) and slightly supercritical ($r_c>0$) values of $Ra$. Here 
$r_c=Ra/Ra_c-1$ is the criticality parameter. The data from these DNS, including $Ra$, measured frequency and wavelength is represented on figure \ref{fig:Dependence_on_Re}.
{We stress that the primary purpose here is not one of 
validation of the growthrate obtained by LSA (even though this comes as a 
by-product), but to answer the question of whether the LSA correctly identifies the mode that ``naturally" emerges from an unstable base state. As such, it is 
essential that the initial condition be unbiased towards a particular mode. 
This is the reason why white noise, rather than the most unstable mode is used 
to perturb the base state in the initial condition.}

\begin{figure}
\begin{center}
\includegraphics[width=\textwidth]{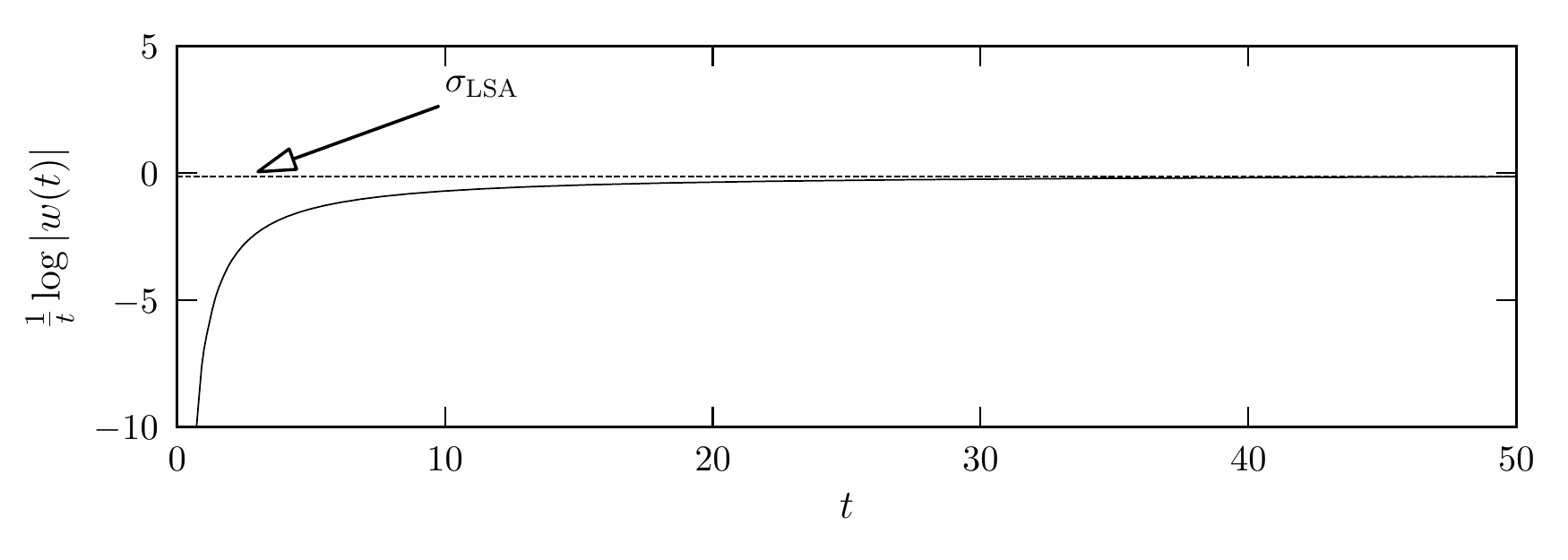}
\end{center}
\caption{Decay rate built on the time history of the $w(t)$ velocity component at $(x,y,z) = (0,0.6,0.5)$ 
obtained by DNS for $Re=200$ and $Ra=10^5$, compared to the growthrate obtained by linear stability 
analysis (LSA).} 
\label{fig:decay}
\end{figure}

In all investigated cases, the perturbation was found to decay for $Ra<Ra_c$ and the flow to bifurcate 
away from the base state for $Ra>Ra_c$. The subcritical decay rate was 
extracted from the DNS with an exponential fit to the long-time decay of $w$ measured at a single point 
in the domain as shown in figure \ref{fig:decay} for $Re=200$ and $Ra=10^5$. The asymptotic decay rate was always found within $0.3\%$ of the prediction of the linear stability analysis, confirming that the fully non-linear 
decay is dominated by the leading mode returned by the linear stability (see Table~\ref{tab:decay} for details).

Further confirmation that the leading mode identified by LSA drives the dynamics near $Ra=Ra_c$ is found 
by comparing critical frequencies and wavelengths, $\omega_c$, $k_c$. In the DNS, the perturbation 
is isolated by subtracting the steady two-dimensional base solution from the result of the 
time-dependent 3D simulation. Again, an agreement with a relative error lower than 
2\% is found between the DNS and the LSA.

\setlength{\tabcolsep}{18pt}
\begin{table}
\begin{center}
\def~{\hphantom{0}}
\begin{tabular}{c c c c c c}
$Re$  & $Ra$   & $r_c$ &  $\sigma$(LSA) & $\sigma$(DNS) & $\epsilon_\sigma(\%)$ \\[2 mm]
$0$   & $5.7\times10^3$ & $-0.0460$ & $-0.0601$& $-0.0602$ & $0.2$\\
$25$   & $2.0\times10^3$ & $-0.1893$ & $-0.1919$& $-0.1917$ & $0.1$\\
$50$   & $6.0\times10^3$ & $-0.0272$ & $-0.0338$& $-0.0337$ & $0.3$\\
$75$   & $1.5\times10^4$ & $-0.2051$ & $-0.1387$& $-0.1385$ & $0.1$\\
$100$   & $3.0\times10^4$ & $-0.1715$ & $-0.1711$& $-0.1712$ & $0.1$\\
$110$   & $4.0\times10^4$ & $-0.0713$ & $-0.1004$& $-0.1005$ & $0.1$\\
$150$   & $7.0\times10^4$ & $-0.0470$ & $-0.1146$& $-0.1145$ & $0.1$\\
$200$   & $1.0\times10^5$ & $ -0.0282$ & $-0.1359$& $-0.1356$ & $0.2$
\end{tabular}
\caption{Comparison of the decay rate computed from the linear stability analysis (LSA) and direct numerical simulation (DNS) for $r_c<0$. $\epsilon_\sigma$ represents the relative error in the computation of growthrate from DNS and LSA.}
\label{tab:decay}
\end{center}
\end{table}

\subsection{Topology and time-dependence of the perturbation near criticality \label{sec:topo}}
\begin{figure}
\begin{center}
\includegraphics[width=\textwidth]{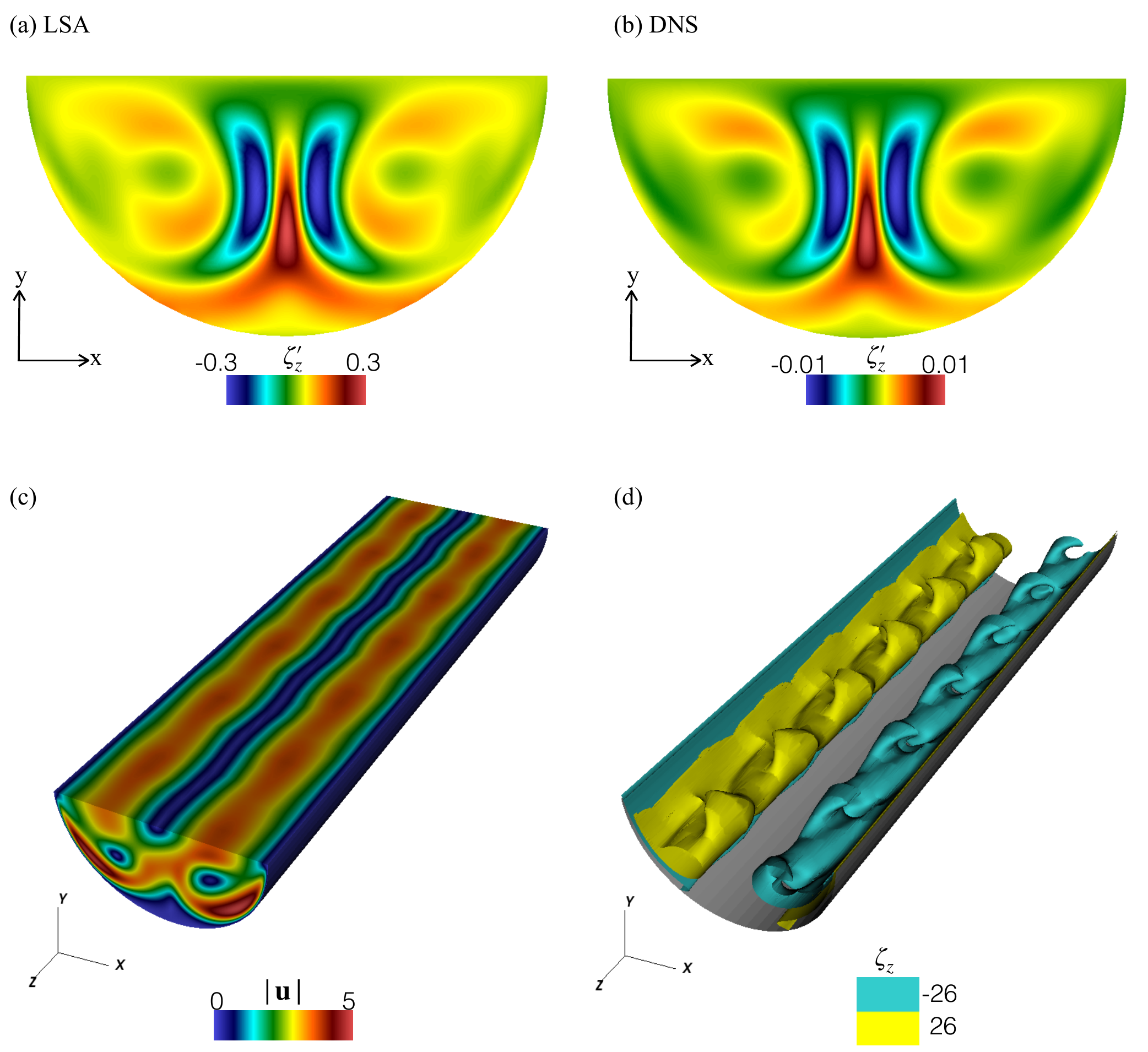}
\end{center}
\caption{For $Re=0$ and $Ra=7 \times 10^3$: (a) Vorticity perturbation along the homogeneous direction $\zeta_z^\prime$ computed from the linear stability analysis (LSA) at $z=0$ plane for $k=6$; (b)  $\zeta_z^\prime$ computed from the direct numerical simulation (DNS) at $z=0$ plane; (c) density plot for the magnitude of the velocity field; and (d) iso-surfaces of the $z$-component of vorticity. A movie 
representing the travelling wave is available in the supplementary material (see supplementary movie 1).} 
\label{fig:DNS_Re_0}
\end{figure}

\begin{figure}
\begin{center}
\includegraphics[width=\textwidth]{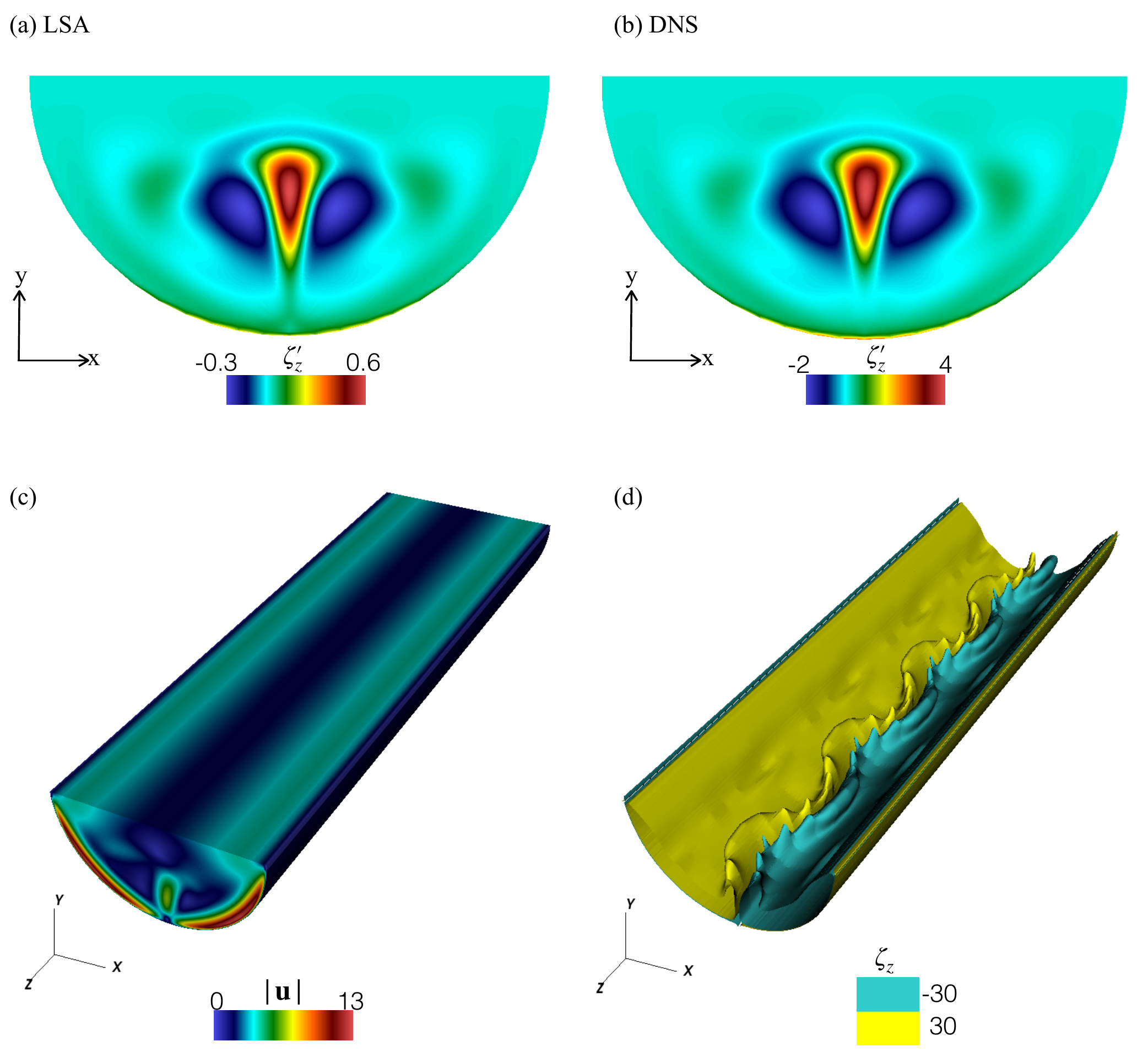}
\end{center}
\caption{For $Re=100$ and $Ra=4 \times 10^4$: (a) Vorticity perturbation along the homogeneous direction $\zeta_z^\prime$ computed from the linear stability analysis (LSA) at $z=0$ plane for $k=4$; (b)  $\zeta_z^\prime$ computed from the direct numerical simulation (DNS) at $z=0$ plane; (c) density plot for the magnitude of the velocity field; and (d) iso-surfaces of the $z$-component of vorticity. 
A movie representing the standing wave is available in the supplementary material (see supplementary movie 2).} 
\label{fig:DNS_Re_100}
\end{figure}

\begin{figure}
\begin{center}
\includegraphics[width=\textwidth]{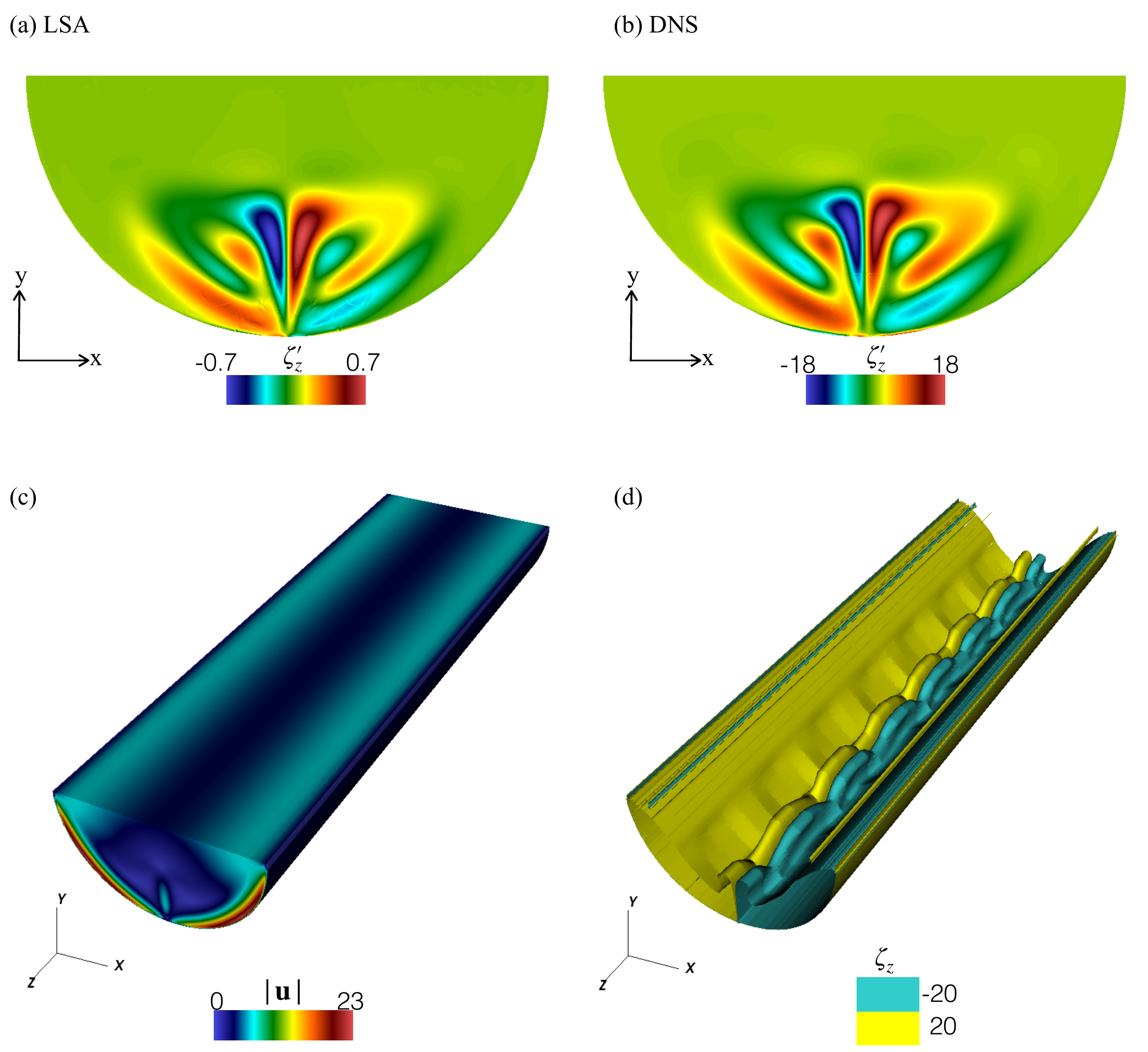}
\end{center}
\caption{For $Re=200$ and $Ra=1.1 \times 10^5$: (a) Vorticity perturbation along the homogeneous direction $\zeta_z^\prime$ computed from the linear stability analysis (LSA) at $z=0$ plane for $k=7$; (b)  $\zeta_z^\prime$ computed from the direct numerical simulation (DNS) at $z=0$ plane; (c) density plot for the magnitude of the velocity field; and (d) iso-surfaces of the $z$-component of vorticity.} 
\label{fig:DNS_Re_200}
\end{figure}

Figures \ref{fig:DNS_Re_0}, \ref{fig:DNS_Re_100} and \ref{fig:DNS_Re_200} respectively 
show the velocity magnitudes and vorticity distributions in weakly supercritical cases associated to 
each of the three instability branches (respectively for $Re=0$, $Re=100$ and $Re=200$).
The time evolution of the perturbation reconstructed from the LSA involves both real and imaginary parts 
as of the LSA eigenvector:
\begin{equation}
{\bf q}^\prime(x,y,z,t)=Re \left\{ \hat{\bf q}(x,y)e^{\sigma t+ i(kz + \omega t)} \right\}.
\end{equation}
The snapshots of the topologies of the perturbation from LSA and DNS  presented on the figures are captured at the same phase.

Unsurprisingly, the topologies of the perturbations found in the DNS and LSA precisely agree too. In all 
cases, the instability originates near the symmetry plane $x=0$, just above the location where 
the jets driven by the baroclinic imbalance on either side of the cavity meet. The driving mechanism is 
a destabilisation of the return jet, with a different behaviour depending on the branch the unstable 
mode belongs to:
modes from Branch II ($Re<75$) develop into a travelling wave along $\mathbf e_z$. Since the travelling wave is made up of two counter-propagative linear waves (respectively associated to each of the 
conjugate eigenvalues), the travelling nature of the wave is determined by the complex amplitude of 
the unstable modes \citep{clune1993_pre}. These cannot be obtained from LSA, but appear in the fully 
non-linear DNS (see figure \ref{fig:DNS_Re_0} and associated animation). By contrast, modes from branch 
I, which are the most unstable for $100\leq Re\leq 110$, always develop into a standing wave with 
transverse oscillations within the $x-y$ plane.

\section{Characterisation of the transition to oscillatory and non-oscillatory states}
\label{sec:ST_model}
\subsection{Stuart--Landau model}
\label{ssec:ST_model}
We now seek to characterise the bifurcation associated to the instabilities identified in the previous section  by means of a truncated Stuart--Landau equation. This model has been widely applied to find the nature of bifurcations in 
a number of fluid flows, among which: flow past a circular cylinder~\citep{Provansal:JFM1987,Dusek:JFM1994,Albarede:JFM1995,Schumm:JFM1994,Henderson:PF1996,Thompson:JFM2004}, staggered cylinder~\citep{Carmo:JFM2008} and rings~\citep{Sheard:JFM2004}, and the flow confined around a 180-degree sharp bend~\citep{Sapardi:JFM2017,pzk2019_jfm}. 
The principle traces back to the equation proposed by 
\citet{Landau:1944} to describe the transition to turbulence, and later used by 
~\citet{Stuart:JFM1958,Stuart:JFM1960} to understand the behaviour of the plane Poiseuille flow. The 
Stuart--Landau model describes the growth and saturation of the complex amplitude $A(t)$ of a 
perturbation near the onset of instability as~\citep{Landaus:book:Fluid} 
\begin{equation}
\frac{dA}{dt} = (\sigma+i\omega)A-l(1+ic)|A|^2A+\mathcal{O}(A^5), \label{eq:Landau_complex}
\end{equation}
where $l\in\mathbb R$ reflects the level of nonlinear saturation and $c\in \mathbb R$ is the Landau 
constant. For $l>0$, the first two terms on the right-hand side of the equation~(\ref{eq:Landau_complex}) provide a good description of the evolution of the perturbation, and the saturation occurs through the 
cubic term. In this case the bifurcation is {\em supercritical}. For $l<0$ , the cubic term accelerates 
the growth of the perturbation, and higher-order terms are needed to saturate the growth. This case 
corresponds to a {\em subcritical} transition. The equations for the time evolution  of the (real) amplitude $|A(t)|$ and phase $\phi(t)$ are obtained 
by substituting $A(t)=|A(t)|e^{i\phi(t)}$ into  equation~(\ref{eq:Landau_complex}), and separating the real and imaginary parts:
\begin{eqnarray}
\frac{d |A|}{dt} & = &\sigma|A| - l |A|^3, \label{eq:Landau_real}\\
\frac{d\phi}{dt} & = &\omega - l c|A|^2. \label{eq:Landau_imag}
\end{eqnarray}
Equation~(\ref{eq:Landau_real}) is rewritten as  
\begin{equation}
\frac{d\log|A|}{dt} = \sigma - l |A|^2.
\label{eq:dlogA_dt}
\end{equation}
As noted by \citet{Sheard:JFM2004}, this form of the Stuart--Landau equation makes it convenient to 
determine $\sigma$ and $l$ from three-dimensional direct numerical simulations (see \S~\ref{ssec:Landau_model_app}), if $A(t)$ is defined as the time-dependent amplitude of one component of 
the velocity perturbation, for example.
The amplitude of the perturbation in the saturated state is readily obtained by setting $\partial_t=0$ 
in ~(\ref{eq:Landau_real}): 
\begin{equation}
|A_{sat}| = \sqrt{\frac{\sigma}{l}}. \label{eq:A_sat}
\end{equation}
Furthermore, if the flow settles down to a time-periodic state with constant amplitude $|A_{sat}|$, 
$d\phi/dt$ becomes the constant angular frequency of oscillation $\omega_{sat}$, and  
equation~(\ref{eq:Landau_imag}) yields, 
\begin{equation}
c = \frac{\omega_{sat}-\omega}{\sigma}. \label{eq:c}
\end{equation}
Thus, the Landau constant $c$ can be determined by computing the oscillation frequency of the perturbation in the linear regime and at the saturation.  
\subsection{Nature of the bifurcations}
\label{ssec:Landau_model_app}
\begin{figure}
\centering
    \begin{tabular}{ll}
    (a) $Re=0$ & (b) $Re=50$\\
    \parbox{0.45\textwidth}{\includegraphics[width=0.49\textwidth]{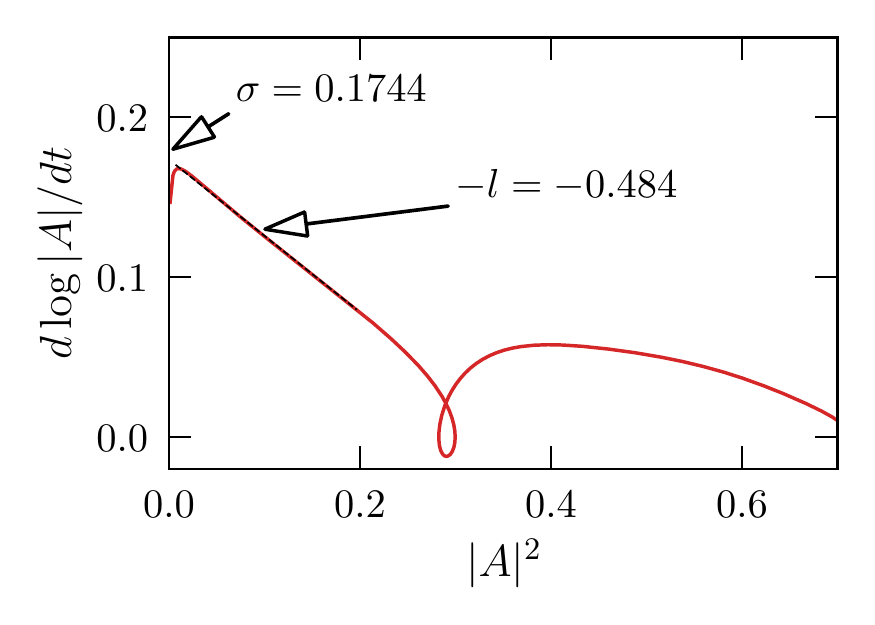}} &
    \parbox{0.5\textwidth}{\includegraphics[width=0.49\textwidth]{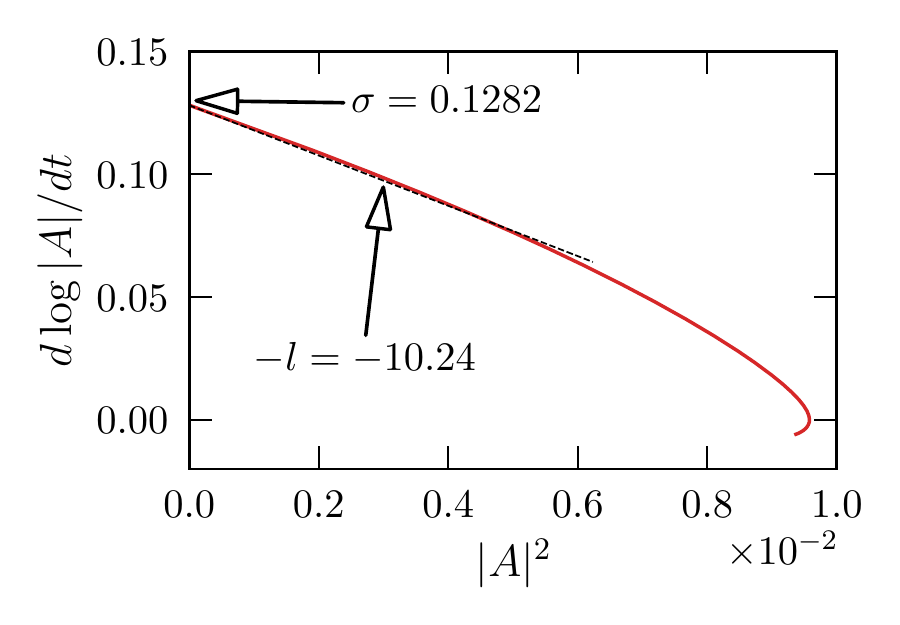}} \\
    (c) $Re=100$ & (d) $Re=200$\\
    \parbox{0.45\textwidth}{\includegraphics[width=0.49\textwidth]{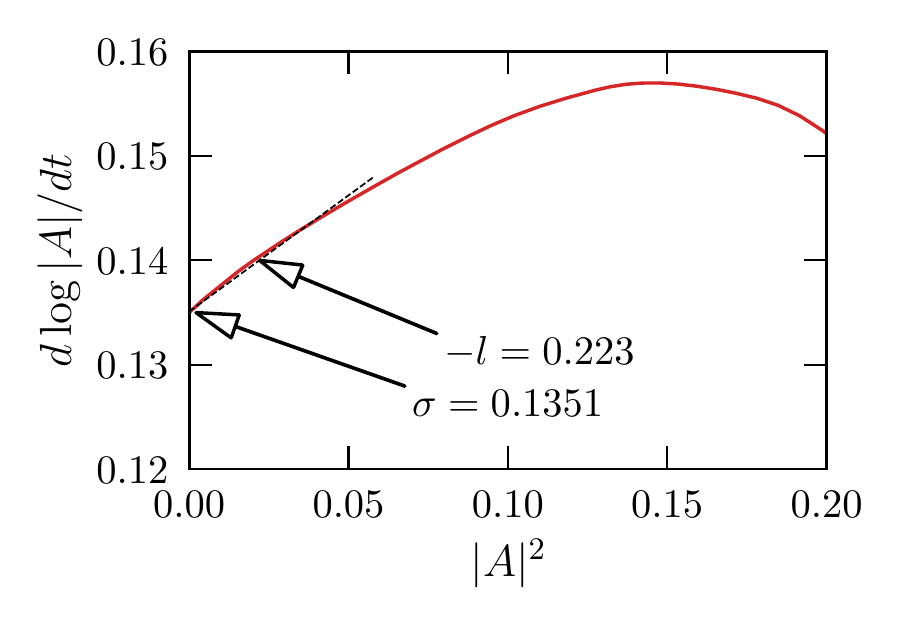}} &
    \parbox{0.5\textwidth}{\includegraphics[width=0.49\textwidth]{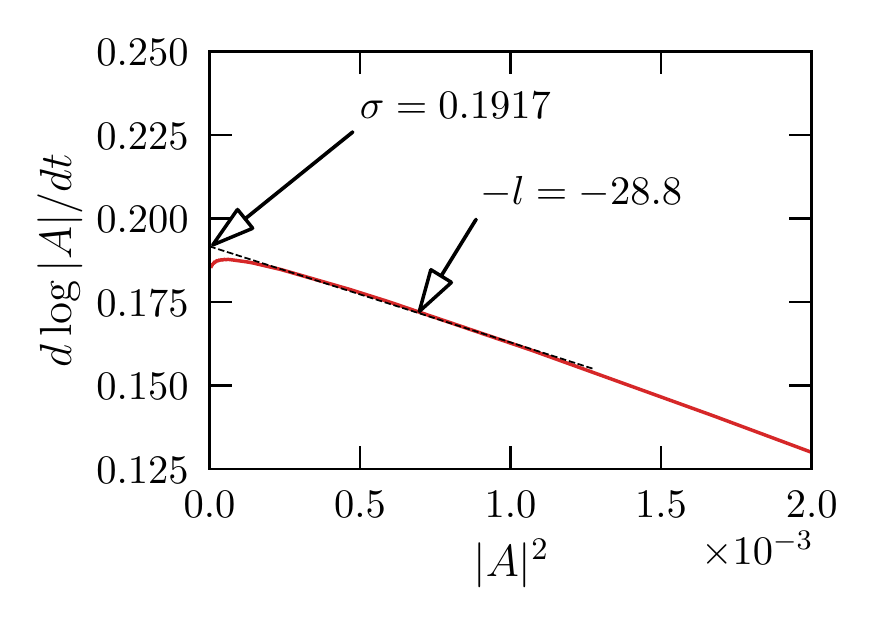}} \\
  \end{tabular}
  \caption{Stuart--Landau analysis: variation of $(d \log|A|/dt)$ vs. $|A|^2$ and extrapolation to $|A|=0$ from which coefficients $\sigma$ and $l$ are obtained for (a) $Re=0$; $Ra=7\times10^3$, (b) $Re=50$; $Ra=7\times10^3$, (c) $Re=100$; $Ra=4\times 10^4$, and (d) $Re=200$; $Ra=1.1\times10^5$. {$|A(t)|$ is obtained from time-series of $w(t)$.} }
\label{fig:Landau_analysis}
\end{figure}

We shall first determine the nature of the bifurcation by calculating constant $l$, and checking its 
sign. We follow \citet{Sheard:JFM2004}, and fit equation~(\ref{eq:dlogA_dt}), to the time 
variation of the envelope of $|A(t)|$ extracted from the signal of  the $z-$component of velocity $w(t)$
obtained at a single location, in the neighbourhood of $|A|=0$. This particular choice for $|A(t)|$ and 
the choice of location itself are guided by the requirement of obtaining a clean enough signal. Other 
choices are possible \citep{Sheard:JFM2004}, based on either local or global variables \citep{pzk2019_jfm}.
The analysis has been carried out in slightly supercritical regimes ($r_c>0$, but small) for all values 
of $Re$ investigated in this paper up to $200$. Four 
representative examples are shown on figure \ref{fig:Landau_analysis}. In all cases, a small area very 
close to $|A|=0$ is dominated by numerical noise. The high precision of our DNS however keeps this 
interval small compared to the area where the linear approximation (\ref{eq:dlogA_dt}) remains valid. 
Fitting of (\ref{eq:dlogA_dt}) in the linear range provides the values of $\sigma$ and $l$. Comparing 
$\sigma$ to the value returned by the linear stability analysis provides mutual validation for the 
linear stability and the DNS, but also provides an estimate for the precision of the fit. Both values 
are reported in table \ref{tab:lsa-dns}. The relative discrepancy remains below $6\%$, except for $Re=0$ where the discrepancy is of $11.6\%$.

\setlength{\tabcolsep}{10pt}
$l$ remains positive at the onset of instability for $0\geq Re \geq 75$, which corresponds to the 
range of values of $Re$ where the instability sets in through branch II modes. In other words, 
branch II modes become unstable through 
a supercritical Hopf bifurcation. For $100\leq Re < 150$, by contrast, $l<0$ indicating that the 
modes of branch I destabilise through a subcritical Hopf bifurcation. This may appear as surprising 
considering the very good agreement between the critical Rayleigh number for the onset of instability 
found by DNS and LSA. Nevertheless, the small value of $l$ suggests that the system may only be 
mildly subcritical.  It is also possible that the addition of white noise of moderate amplitude may not 
suffice to drive the growth of subcritical perturbations. {Indeed, we verified that 
further increasing the standard deviation of the added white noise to $0.01$ and $0.1$ did not alter 
the results}. Whether transition could occur for $Ra<Ra_c$, could be 
answered by analysing whether non-modal perturbations could grow \citep{schmid01}, and whether 
{optimal} perturbation of sufficient amplitudes could trigger a subcritical transition 
to {another state, as it does for turbulence in pipes}~\citep{Pringle:JFM2012}. For 
$Re\geq150$, 
the instability occurs through non-oscillatory modes from branch III.  $l$ becomes positive again, 
indicating that this transition is of supercritical  type { (supercritical pitchfork)}. In contrast with bifurcation through 
branch I, the supercritical nature of the instability of modes from branches II and III is consistent 
with the excellent agreement between LSA and DNS on the value of $Ra$ for the onset of instability and 
further establishes the relevance of LSA to determine the stability of the flow in these cases.
\setlength{\tabcolsep}{5pt}
\begin{table}
\begin{center}
\def~{\hphantom{0}}
\begin{tabular}{c c c | c c c | c c c c}
$Re$  & $Ra$  & $r_c$ & $\sigma$(LSA) & $\sigma$(DNS) & $\epsilon_\sigma(\%)$  & $\omega$(LSA) & $\omega$(DNS) & $\epsilon_\omega(\%)$ & $\omega-\omega_{sat}$\\[2 mm]
$0$   & $7.0 \times 10^3$ & $0.1715$ & $0.1541$ & $0.1744$ & $11.6$ &  $5.9690$ & $5.8698$ & $1.7$ & $0$\\
$25$   & $3.0 \times 10^3$ & $0.2161$ & $0.0731$ & $0.0736$ & $0.7$ & $3.3498$ & $3.3211$ & $0.9$ & $0$\\
$50$   & $7.0 \times 10^3$ & $0.1348$ & $0.1209$ & $0.1282$ & $5.7$ & $4.3140$ & $4.3982$ & $1.9$ & $0$\\
$75$   & $2.0 \times 10^4$ & $0.0599$ & $0.0711$ & $0.0730$ & $2.7$ & $5.4808$ & $5.5099$ & $0.5$ & $0$\\
$100$  & $4.0 \times 10^4$ & $0.1047$ & $0.1344$ & $0.1351$ & $0.5$ & $11.599$ & $11.624$ & $0.2$ & $0.942$\\
$110$  & $4.5 \times 10^4$ & $0.0448$ & $0.0607$ & $0.0622$ & $2.4$ & $11.890$ & $11.938$ & $0.4$ & $0.628$\\
$150$  & $8.0 \times 10^4$  & $0.0892$ & $0.2100$ & $0.2084$ & $0.8$ & $0$ & $0$ &--- & ---\\
$200$  & $1.1 \times 10^5$ & 0.0690 & $0.1873$ & $0.1917$ & $2.3$ & $0$ & $0$ & --- & --- \\
\end{tabular}
\caption{Comparison of the growthrate and frequency  computed from the linear stability analysis (LSA) and direct numerical simulation (DNS) for $r_c>0$. $\epsilon_\omega$ represents the relative error in the computation of frequency from DNS and LSA.}
\label{tab:lsa-dns}
\end{center}
\end{table}

\subsection{Saturated states}
\label{sec:sat_states}
\begin{figure}
\begin{center}
\includegraphics[scale = 0.65]{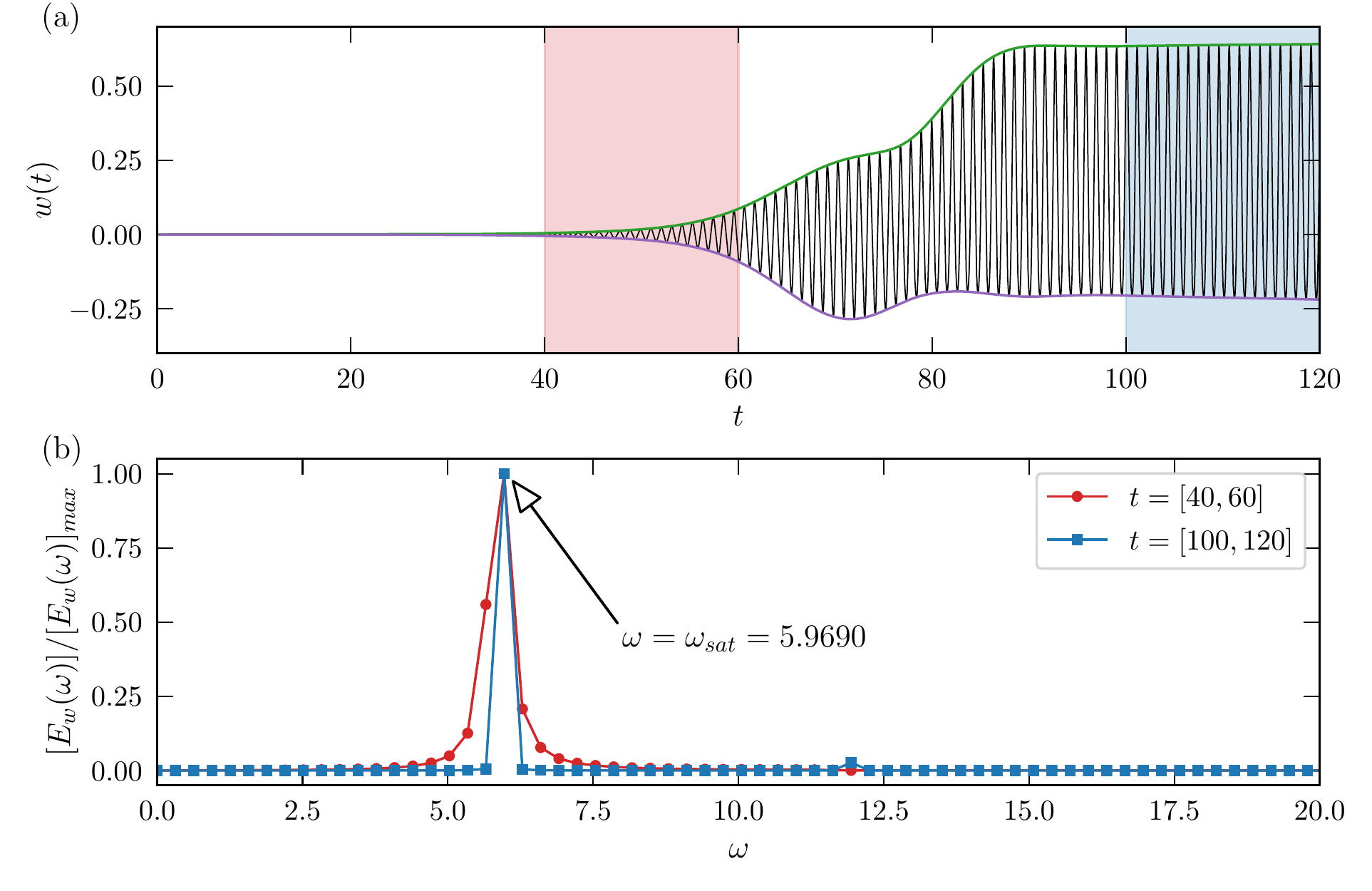}
\end{center}
\caption{(a) Time history of $w(t)$ velocity component measured at $(x,y,z)=(-0.6,0.6,0.5)$ for $Re=0$ and $Ra=7 \times 10^3$. (b) Frequency spectra obtained from the time series of $w(t)$ respectively near the onset and near saturation.}  
\label{fig:time_Re_0}
\end{figure}

\begin{figure}
\begin{center}
\includegraphics[scale = 0.65]{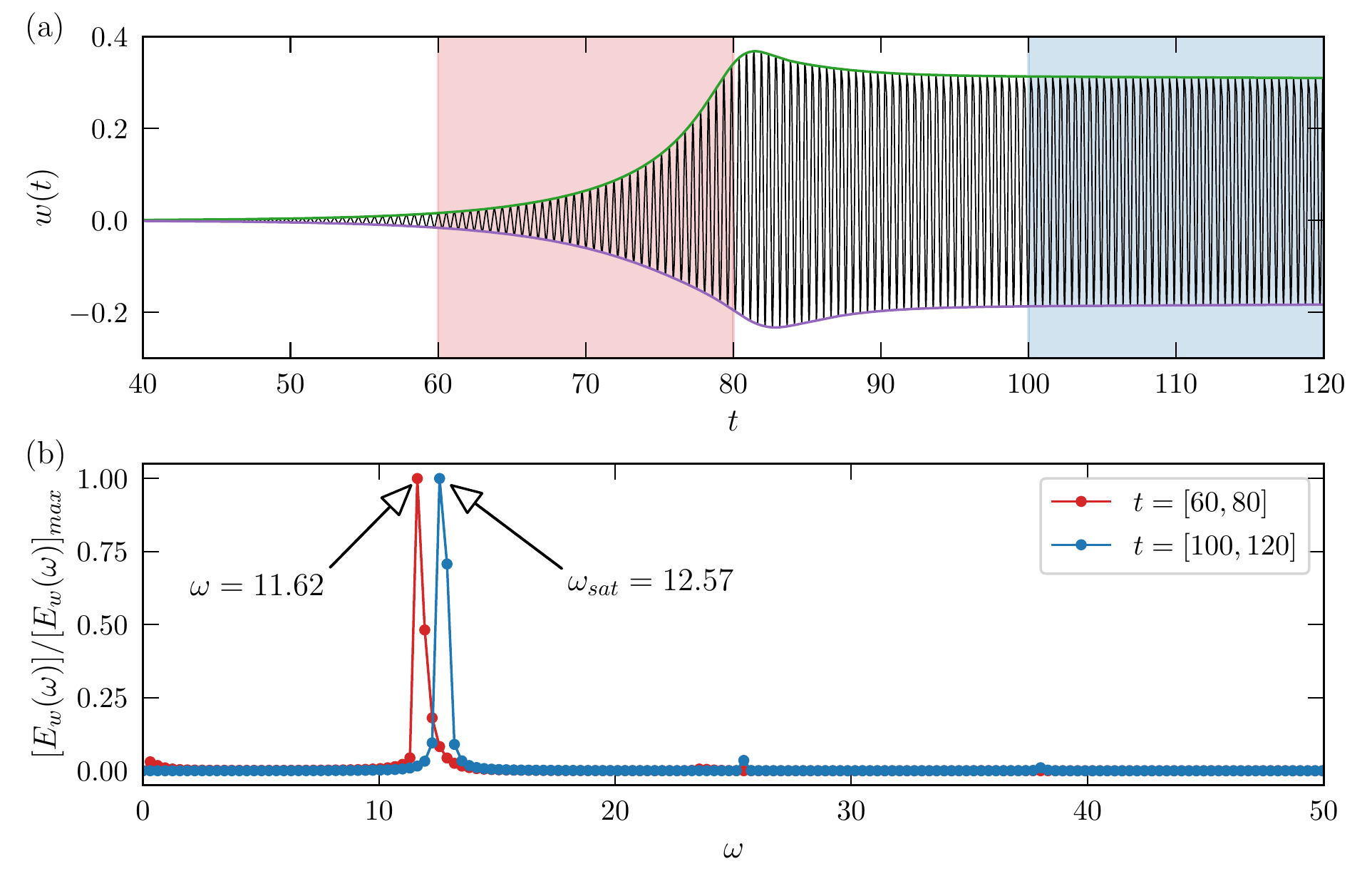}
\end{center}
\caption{(a) Time history of $w(t)$ velocity component measured at $(x,y,z)=(-0.6,0.6,0.5)$ for $Re=100$ and $Ra=4 \times 10^4$. (b) Frequency spectra obtained from the time series of $w(t)$ respectively near the onset and near saturation.}  
\label{fig:time_Re_100}
\end{figure}

Finally, we shall characterise the saturated states. The analysis was carried out for each investigated 
value of $Re$ up to $Re=200$, of which we present three typical cases. 
As discussed in \S~\ref{ssec:nonlinear_analysis}, the three-dimensional DNS were performed for two values 
of $Ra$ in each case, one slightly subcritical and one slightly supercritical. As an example, figure~\ref{fig:time_Re_0}(a) shows the time history of $w$ at a single 
point of the domain for the weakly supercritical case of $Re=0$ and $Ra=7\times 10^3$.  
The normalised frequency spectrum of the time-series 
\begin{equation}
E_w(\omega)=\frac{1}{2}|\hat{w}(\omega)|^2,
\end{equation}
where $\hat{w}(\omega)$ is the Fourier transform of $w(t)$, is then calculated over two intervals: one 
near the onset ($40\leq t \leq 60$), and one in the saturated regime ($100\leq t \leq 120$). Both are 
represented in figure~\ref{fig:time_Re_0}(b). The initial and saturated frequencies are nearly identical (with relative error to the numerical precision)
and differ by $1.7\%$ from the frequency returned by the linear stability analysis. Given that the 
discrepancy between the frequencies near the onset and in the saturated state differ by much less than 
the error between the frequency near the onset predicted by LSA and DNS, we consider them to be 
identical. In this case, from equation~(\ref{eq:c}), the Landau constant $c$ is zero, to the precision 
of our simulations. 
Following this approach, non-zero values of $c$ were found for other values of $Re$ when 
$\omega-\omega_{sat}$ significantly 
exceeded the discrepancy between LSA and DNS. One such example is shown in figure \ref{fig:time_Re_100}, for $Re=100$, where $c=-7.03\pm0.01$ and the discrepancy between LSA and DNS 
frequencies at onset is $0.2\%$.

\begin{figure}
\begin{center}
\includegraphics[scale = 0.65]{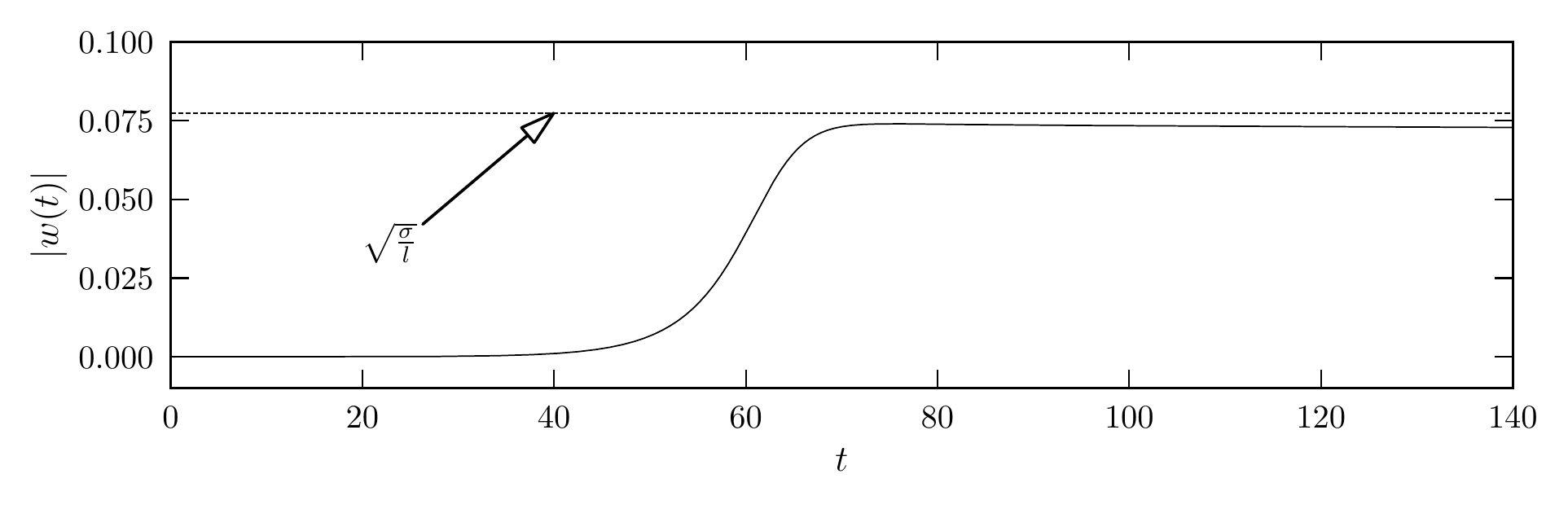}
\end{center}
\caption{Time history of $w(t)$ velocity component measured at $(x,y,z)=(-0.5,0.6,0.5)$ for $Re=200$ and $Ra=1.1 \times 10^5$.}  
\label{fig:time_Re_200}
\end{figure}

All calculated values of $c$ are reported in table \ref{tab:summary}. Here again, the three branches 
identified in  \S~\ref{sec:lsa} exhibit different behaviours: when the instability arises out of 
 modes in branch II, the Landau constant is zero. A shift in frequency does appear as soon as the instability is due to modes belonging to branch I, leading to a negative Landau constant. Finally, as branch 
III becomes dominant, DNS confirm that the steady saturated state of the mode predicted by the linear 
stability is non-oscillatory. In this case, the saturated amplitude predicted by equation (\ref{eq:A_sat}) 
matches closely that observed in the DNS, as shown on figure~\ref{fig:time_Re_200}.
\setlength{\tabcolsep}{9pt}
\begin{table}
\begin{center}
\def~{\hphantom{0}}
\begin{tabular}{c c c c c c}
$Re$  & $Ra$  & $l$ & $c$ & Branch& Nature of transition \\[2 mm]
$0$   & $7 \times 10^3$ & $0.484\pm0.001$ & $0$& II & Supercritical Hopf \\
$25$   & $3 \times 10^3$ & $1.475\pm0.001$ & $0$ & II&Supercritical Hopf \\
$50$   & $7 \times 10^3$ & $10.24\pm0.03$ & $0$ & II& Supercritical Hopf \\
$75$   & $2 \times 10^4$ & $2.010\pm0.001$ & $0$ & II& Supercritical Hopf \\
$100$   & $4 \times 10^4$ & $-0.223\pm0.001$ & $-7.03\pm0.01$ & I &Subcritical Hopf \\
$110$   & $4.5 \times 10^4$ & $-10.2\pm0.1$ & $-0.04\pm0.01$ & I & Subcritical Hopf \\
$150$   & $8 \times 10^4$ & $1.8470\pm0.0001$ & --- & III &Supercritical pitchfork \\
$200$   & $1.1 \times 10^5$ & $28.8\pm0.1$ & --- & III & Supercritical pitchfork
\end{tabular}
\caption{Summary of Stuart--Landau analysis of three-dimensional direct numerical simulation for $Re=0$ to $Re=200$ at $Ra>Ra_c$.}
\label{tab:summary}
\end{center}
\end{table}

{
\subsection{Heat flux in the saturated states}
To finish, we compare the  Nusselt number  $Nu$ in base two-dimensional state and the bifurcated 
three-dimensional state at the onset of instability for all Reynolds numbers in figure \ref{fig:Nu}(b). 
The maximum difference between them is $0.2\%$. Thus, there is hardly any change in the heat transfer.  
This can be understood as the boundary conditions impose that the horizontal heat flux must be conserved between the two boundaries with 
periodic boundary conditions. This precludes any horizontal redirection of the horizontal heat flux. 
Furthermore, the time and spatially periodic character of the instability implies that a drastic 
change would have had to take place between the streamlines in the $(x,y)$ plane of the base flow 
and those of the bifurcated states for a significant change in $Nu$ to be observed. Nevertheless, 
\ref{fig:Nu}(b) expresses that heat transfer at the onset of the instability is all the lower as 
$Re$ is high because of the suppression of the convection by the through-flow.}

\section{Conclusions}
\label{sec:conclusion}
We presented a systematic analysis of the mixed baroclinic convection in a 
pool with hot homogeneous through-flow fed  in at the upper boundary and escaping 
through a porous, semi-circular, cold isothermal lower boundary. Linear stability analysis, DNS and bifurcation analysis have brought answers to the 4 questions set out in introduction:
\begin{enumerate}
\item 
The base flow is driven by a baroclinic imbalance along the 
lower boundary, that peaks at its corners. Downward flows on either 
side of the pool meet in the symmetry plane to form two two-dimensional
 counter-rotating rolls. 
Being deprived of shear, the sole effect of the through-flow on the base flow 
is to displace the convective rolls downward. Once these are confined by the 
lower boundary, further increasing the through-flow (\emph{i.e.} $Re$) leads to their progressive 
suppression and their eventual disappearance for $Re\geq200$. {Interestingly, at low 
through-flow, this type of convection is less effective at carrying the heat downward as conduction 
in solid moving downwards at the same velocity.}
\item 
A consequence of the suppression of the rolls is the stabilisation of the flow to infinitesimal disturbances. Indeed, the critical Rayleigh for linear stability of the base flow $Ra_c$ first decreases as 
the rolls are stretched down ($Re\leq25$) to then increase with $Re$ as they become suppressed.
\item  
The base flow was found susceptible to three distinct types of infinitesimal 
perturbations, all of them with maximum vorticity near the symmetry plane, 
where the rolls meet.  For $Re\leq75$ the most unstable mode (type II) is a wave travelling in the 
$\mathbf e_z $ direction. For $100\leq Re\leq 110$, instability sets in as a standing oscillation (type 
I mode), whereas for $Re\geq150$ the dominating mode is non-oscillating (type III). 
DNS have confirmed the findings of the linear stability analysis, both in terms of topology of the modes, and the critical parameters at the onset (critical Rayleigh number, wavelength, and frequencies). 
\item Most interestingly Stuart--Landau analysis conducted on DNS data revealed that the nature of the 
bifurcation associated to the three modes varies too. While mode II and III appear at a supercritical bifurcation, the onset of mode I is subcritical. 
\end{enumerate}

Nevertheless, the values of the constant $l$ indicate a low level of subcriticality. This may partly 
explain why LSA and DNS are still in agreement at the onset of the subcritical branch. Still, the 
change of nature of the bifurcation near the onset is an interesting feature of this problem. It raises 
the question of whether the system would support other convective states located on subcritical branches 
not connected to the base state considered in this study. These would need to be ignited from a 
different set of initial conditions. Such phenomenology was recently found in numerical models for 
rotating convection in the Earth's core, {where the curvature of the boundaries plays 
an important role too} \citep{Guervilly:JFM2016}.

These results introduce a number of new features, compared to the reference cases of convection 
in an inclined channel \citep{gage1968_jfm} and mixed convection in a cavity \citep{papanicolaou1992_jfm}, despite the similarities pointed out in introduction. 
Unlike convection in an inclined channel, longitudinal rolls are present in the base flow because the 
close, semicircular shape of the boundary does not support an open flow in the direction of the 
baroclinic jets. As such, all unstable modes 
involve a form of longitudinal variation. As direct consequence, the transversal travelling waves found 
in inclined channels cannot exist but remarkably, longitudinal travelling waves exist instead. 
Furthermore, while travelling waves found in the inclined channel 
problem are always subject to a secondary instability, they evolve into a stable periodic flow in our 
semi-cylindrical geometry. Unlike the cavity flow where 
the through-flow is located on one side \citep{papanicolaou1992_jfm}, the homogeneous through-flow 
studied here suppresses wave propagation, which turns into standing oscillations for $Re\geq100$ and finally into a non-oscillatory unstable mode for $Re\geq150$.

Finally, the relevance of these findings to continuous casting is again partial: we do not submit the 
phenomenology found here as a full explanation of the dynamics of these processes. {Nevertheless, 
evidence of oscillatory phenomena in continuous casting processes \citep{Dorward:Al} suggest that the physical mechanisms 
involved play a role amongst other effects and the phenomena 
we described may be observed in some form, in specific configurations where they are not overshadowed by 
other mechanisms not considered here (such as, double diffusion or variations of the pool shape with the flow parameters).}

\section*{Acknowledgements}
This research was funded by Constellium Technology Center (C-TEC). We thank  H. M. Blackburn and D. Moxey  for helping in adapting the codes DOG and Nektar++ respectively. A. Poth\'erat acknowledges support from the Royal Society under the Wolfson Research Merit Award Scheme (Grant reference WM140032).


\begin{thebibliography}{10}

\bibitem{Albarede:JFM1995}
P.~Albar{\`e}de and M.~Provansal.
\newblock {Quasi-periodic cylinder wakes and the Ginzburg{\textendash}Landau
  model}.
\newblock {\em J. Fluid Mech.}, 291:191--222, 1995.

\bibitem{aujogue2015_pf}
K.~Aujogue, A.~Poth\'erat, and B.~Sreenivasan.
\newblock Onset of plane layer magnetoconvection at low {Ekman} number.
\newblock {\em Phys. Fluids}, 27(10):106602, 2015.

\bibitem{Barkley:IJNMF2008}
D.~Barkley, H.~M. Blackburn, and S.~J. Sherwin.
\newblock {Direct optimal growth analysis for timesteppers}.
\newblock {\em Intl J. Numer. Meth. Fluids}, 57(9):1435--1458, July 2008.

\bibitem{Barkley:JFM2002}
D.~Barkley, M.~G.~M. Gomes, and R.~D. Henderson.
\newblock {Three-dimensional instability in flow over a backward-facing step}.
\newblock {\em J. Fluid Mech.}, 473:167--190, 2002.

\bibitem{Barkley:JFM1996}
D.~Barkley and R.~D. Henderson.
\newblock {Three-dimensional Floquet stability analysis of the wake of a
  circular cylinder}.
\newblock {\em J. Fluid Mech.}, 322:215--241, 1996.

\bibitem{Blackburn:JFM2004}
H.~M. Blackburn and S.~J. Sherwin.
\newblock {Formulation of a Galerkin spectral element{\textendash}Fourier
  method for three-dimensional incompressible flows in cylindrical geometries}.
\newblock {\em Journal of Computational Physics}, 197(2):759--778, July 2004.

\bibitem{briggs1985_jht}
D.~Briggs and D.~Jones.
\newblock Two-dimensional periodic natural convection in a rectangular
  enclosure of aspect ratio one.
\newblock {\em ASME. J. Heat Transfer}, 107(4):850--854, 1985.

\bibitem{Cantwell:CPC2015}
C.~D. Cantwell, D.~Moxey, A.~Comerford, A.~Bolis, G.~Rocco, G.~Mengaldo, D.~D.
  Grazia, S.~Yakovlev, J.~E. Lombard, D.~Ekelschot, B.~Jordi, H.~Xu,
  Y.~Mohamied, C.~Eskilsson, B.~Nelson, P.~Vos, C.~Biotto, R.~M. Kirby, and
  S.~J. Sherwin.
\newblock {Nektar++: An open-source spectral/hp element framework}.
\newblock {\em Comput. Phys. Commun.}, 192:205--219, July 2015.

\bibitem{Canuto:book:SpectralFluid}
C.~Canuto, M.~Y. Hussaini, A.~Quarteroni, and T.~A. Zhang.
\newblock {\em {Spectral Methods in Fluid Turbulence}}.
\newblock Springer, Berlin, 1988.

\bibitem{Carmo:JFM2008}
B.~S. Carmo, S.~J. Sherwin, P.~W. Bearman, and R.~H.~J. Willden.
\newblock {Wake transition in the flow around two circular cylinders in
  staggered arrangements}.
\newblock {\em J. Fluid Mech.}, 597:1--29, 2008.

\bibitem{Chandrasekhar:book}
S.~Chandrasekhar.
\newblock {\em {Hydrodynamic and Hydromagnetic Stability}}.
\newblock Clarendon Press, 1968.

\bibitem{Clever:JFM1974}
R.~M. Clever and F.~H. Busse.
\newblock {Transition to time-dependent convection}.
\newblock {\em J. Fluid Mech.}, 65(04):625--21, 1974.

\bibitem{clune1993_pre}
T.~Clune and E.~Knobloch.
\newblock Pattern selection in rotating convection with experimental boundary
  conditions.
\newblock {\em Phys. Rev. E}, 47(4):2536--2540, 1993.

\bibitem{Dorward:Al}
R.~Dorward, D.~Beerntsen, and K.~Brwon.
\newblock {Banded segregation patterns in DC-cast Al-Zn-Mg-Cu alloy ingots and
  their effect on plate properties}.
\newblock {\em Aluminium}, 72(4):251--259, 1996.

\bibitem{Dusek:JFM1994}
J.~Du{\v s}ek, P.~Le~Gal, and P.~Frauni{\'e}.
\newblock {A numerical and theoretical study of the first Hopf bifurcation in a
  cylinder wake}.
\newblock {\em J. Fluid Mech.}, 264:59--80, 1994.

\bibitem{Flood:MST1994}
S.~C. Flood and P.~A. Davidson.
\newblock {Natural convection in aluminium direct chill cast ingot}.
\newblock {\em Materials Science and Technology}, 10(8):741--752, 1994.

\bibitem{fujimura1993_jfm}
K.~Fujimura and R.~E. Kelly.
\newblock Mixed mode convection in an inclined slot.
\newblock {\em J. Fluid Mech.}, 246:545--568, 1993.

\bibitem{fujimura1995_pf}
K.~Fujimura and R.~E. Kelly.
\newblock Interaction between longitudinal convection rolls and transverse
  waves in unstably stratified plane {Poiseuille} flow.
\newblock {\em Phys. Fluids}, 7(1):68--79, 1995.

\bibitem{gage1968_jfm}
K.~S. Gage and W.~H. Reid.
\newblock The stability of thermally stratified plane {Poiseuille} flow.
\newblock {\em J. Fluid Mech.}, 33(1):21--32, 1968.

\bibitem{GeuzaineIJNME:2009}
C.~Geuzaine and J.-F. Remacle.
\newblock {Gmsh Reference Manual. Gmsh: A Three-Dimensional Finite Element Mesh
  Generator With Built-in Pre-and Post-Processing Facilities}.
\newblock {\em Int. J. Numer. Methods Eng.}, 79(1309), 2009.

\bibitem{Guervilly:JFM2016}
C.~Guervilly and P.~Cardin.
\newblock {Subcritical convection of liquid metals in a rotating sphere using a
  quasi-geostrophic model}.
\newblock {\em J. Fluid Mech.}, 808:61--89, Oct. 2016.

\bibitem{hart1971_jfm}
J.~E. Hart.
\newblock Stability of the flow in a differentially heated inclined box.
\newblock {\em J. Fluid Mech.}, 47(3):547--576, 1971.

\bibitem{hart1979_arfm}
J.~E. Hart.
\newblock Finite amplitude baroclinic instability.
\newblock {\em Ann. Rev. Fluid Mech.}, 11(1):147--172, 1979.

\bibitem{Henderson:PF1996}
R.~D. Henderson and D.~Barkley.
\newblock {Secondary instability in the wake of a circular cylinder}.
\newblock {\em Phys. Fluids}, 8(6):1683--1685, 1996.

\bibitem{jaluria1980}
Y.~Jaluria.
\newblock {\em Natural Convection Heat and Mass Transfer}.
\newblock Pergamon, 1980.

\bibitem{james1987_jas}
I.~N. James.
\newblock Suppression of baroclinic instability in horizontally sheared flows.
\newblock {\em J. Atmos. Sci.}, 44(24):3710--3720, 1987.

\bibitem{scipy}
E.~Jones, T.~Oliphant, P.~Peterson, and {\em et al.}
\newblock {SciPy}: Open source scientific tools for {Python}, 2019.

\bibitem{karniadakis_spectral:book}
G.~Karniadakis and S.~J. Sherwin.
\newblock {\em Spectral/hp Element Methods for CFD}.
\newblock Oxford University Press, 1999.

\bibitem{Karniadakis:JCP1991}
G.~E. Karniadakis and S.~A.~O. Israeli, M.~and.
\newblock {High-order splitting methods for the incompressible Navier--Stokes
  equations}.
\newblock {\em J. Comput. Phys.}, 97(2):414--443, Dec. 1991.

\bibitem{kelly1994_aam}
R.~Kelly.
\newblock The onset and development of thermal convection in fully developed
  shear flows.
\newblock volume~31 of {\em Advances in Applied Mechanics}, pages 35 -- 112.
  Elsevier, 1994.

\bibitem{korpela1974_ijhmt}
S.~A. Korpela.
\newblock A study on the effect of {Prandtl} number on the stability of the
  conduction regime of natural convection in an inclined slot.
\newblock {\em Int. J. Heat Mass Transf.}, 17(2):215 -- 222, 1974.

\bibitem{kuznetsov1997_cht}
A.~V. Kuznetsov.
\newblock Double-diffusive convection during continuous strip casting.
\newblock {\em CHT'97 - Advances in Computational Heat Transfer. Proceedings of
  the International Symposium - \c Cesme, Turkey, May 26 - 30, 1997}, 1997.

\bibitem{Landau:1944}
L.~D. Landau.
\newblock {On the problem of turbulence}.
\newblock {\em C. R. Acad. Sci. UESS}, 44(311), 1944.

\bibitem{Landaus:book:Fluid}
L.~D. Landau and E.~M. Lifsitz.
\newblock {\em {Fluid Mechanics}}.
\newblock Pergamon Press, Oxford, 1987.

\bibitem{Mao:PF2014}
X.~Mao and H.~M. Blackburn.
\newblock {The structure of primary instability modes in the steady wake and
  separation bubble of a square cylinder}.
\newblock {\em Phys. Fluids}, 26(7):074103--10, 2014.

\bibitem{nakagawa1957_prsa}
Y.~Nakagawa.
\newblock Experiments on the instability of a layer of mercury heated from
  below and subject to the simultaneous action of a magnetic field and
  rotation.
\newblock {\em Proc. R. Soc. A}, 242(1228):81--88, 1957.

\bibitem{nicolas2000_ijhmt}
X.~Nicolas, J.-M. Luijkx, and J.-K. Platten.
\newblock Linear stability of mixed convection flows in horizontal rectangular
  channels of finite transversal extension heated from below.
\newblock {\em Int. J. Heat Mass Transf.}, 43(4):589 -- 610, 2000.

\bibitem{papanicolaou1992_jfm}
E.~Papanicolaou and Y.~Jaluria.
\newblock Transition to a periodic regime in mixed convection in a square
  cavity.
\newblock {\em J. Fluid Mech.}, 239:489--509, 1992.

\bibitem{pierrehumbert1995_arfm}
R.~T. Pierrehumbert and K.~L. Swanson.
\newblock Baroclinic instability.
\newblock {\em Ann. Rev. Fluid Mech.}, 27(1):419--467, 1995.

\bibitem{Pitz:JFM2017}
D.~B. Pitz, O.~Marxen, and J.~W. Chew.
\newblock {Onset of convection induced by centrifugal buoyancy in a rotating
  cavity}.
\newblock {\em J. Fluid Mech.}, 826:484--502, Aug. 2017.

\bibitem{pzk2019_jfm}
A.~Poth\'erat and L.~Zhang.
\newblock Dean flow and vortex shedding in a three-dimensional 180$^o$ sharp
  bend.
\newblock {\em ArXiv reprint}, page arXiv:1807.10950, 2018.

\bibitem{Pringle:JFM2012}
C.~C.~T. Pringle, A.~P. Willis, and R.~R. Kerswell.
\newblock {Minimal seeds for shear flow turbulence: using nonlinear transient
  growth to touch the edge of chaos}.
\newblock {\em J. Fluid Mech.}, 702:415--443, May 2012.

\bibitem{Provansal:JFM1987}
M.~Provansal, C.~Mathis, and L.~Boyer.
\newblock {B{\'e}nard-von K{\'a}rm{\'a}n instability: transient and forced
  regimes}.
\newblock {\em J. Fluid Mech.}, 182:1--22, 1987.

\bibitem{Sapardi:JFM2017}
A.~M. Sapardi, W.~K. Hussam, A.~Poth{\'e}rat, and G.~J. Sheard.
\newblock {Linear stability of confined flow around a 180-degree sharp bend}.
\newblock {\em J. Fluid Mech.}, 822:813--847, 2017.

\bibitem{schmid01}
P.~J. Schmid and D.~S. Henningson.
\newblock {\em Stability and Transition in Shear Flows}.
\newblock Spinger-Verlag New York, 2001.

\bibitem{Schumm:JFM1994}
M.~Schumm, E.~Berger, and P.~A. Monkewitz.
\newblock {Self-excited oscillations in the wake of two-dimensional bluff
  bodies and their control}.
\newblock {\em J. Fluid Mech.}, 271:17--53, 1994.

\bibitem{Sheard:JFM2004}
G.~J. Sheard, M.~C. Thompson, and K.~Hourigan.
\newblock {From spheres to circular cylinders: non-axisymmetric transitions in
  the flow past rings}.
\newblock {\em J. Fluid Mech.}, 506:45--78, 2004.

\bibitem{sheng2000_mmtb}
D.~Sheng and L.~Jonsson.
\newblock Two-fluid simulation on the mixed convection flow pattern in a
  nonisothermal water model of continuous casting tundish.
\newblock {\em Metallurgical and Materials Transactions B}, 31(4):867--875,
  2000.

\bibitem{shome1995_ijhmt}
B.~Shome and M.~Jensen.
\newblock Mixed convection laminar flow and heat transfer of liquids in
  isothermal horizontal circular ducts.
\newblock {\em Int. J. Heat Mass Transf.}, 38(11):1945 -- 1956, 1995.

\bibitem{Stuart:JFM1958}
J.~T. Stuart.
\newblock {On the non-linear mechanics of hydrodynamic stability}.
\newblock {\em J. Fluid Mech.}, 4(1):1--21, 1958.

\bibitem{Stuart:JFM1960}
J.~T. Stuart.
\newblock {On the non-linear mechanics of disturbances in parallel flows, Part
  1. The basic behavior in plane Poiseuille flow}.
\newblock {\em J. Fluid Mech.}, 9(353--370), 1960.

\bibitem{brian2001_isiji}
B.~G. Thomas and L.~Zhang.
\newblock Mathematical modeling of fluid flow in continuous casting.
\newblock {\em ISIJ International}, 41(10):1181--1193, 2001.

\bibitem{Thomas:PRE1998}
L.~Thomas, W.~Pesch, and G.~Ahlers.
\newblock {Rayleigh-B{\'e}nard convection in a homeotropically aligned nematic
  liquid crystal}.
\newblock {\em Phys. Rev. E}, 58(5):5885--5897, 1998.

\bibitem{Thompson:JFM2004}
M.~C. Thompson and P.~Le~Gal.
\newblock {The Stuart{\textendash}Landau model applied to wake transition
  revisited}.
\newblock {\em Eur. J. Mech. B/Fluids}, 23(1):219--228, 2004.

\bibitem{Tritton:book}
D.~J. Tritton.
\newblock {\em {Physical Fluid Dynamics}}.
\newblock Clarendon Press, Oxford, 1988.

\bibitem{Tuckerman:Springer2000}
L.~S. Tuckerman and D.~Barkley.
\newblock {Bifurcation Analysis for Timesteppers}.
\newblock In {\em Numerical Methods for Bifurcation Problems and Large-Scale
  Dynamical Systems}, pages 453--466. Springer, 2000.

\bibitem{Vallis:Book}
G.~K. {Vallis}.
\newblock {\em Atmospheric and Oceanic Fluid Dynamics}.
\newblock Cambridge University Press, Cambridge, 2006.

\bibitem{vo2017_prf}
T.~Vo, A.~Poth\'erat, and G.~J. Sheard.
\newblock Linear stability of horizontal, laminar fully developed,
  quasi-two-dimensional liquid metal duct flow under a transverse magnetic
  field and heated from below.
\newblock {\em Phys. Rev. Fluids}, 2(3):033902, 2017.

\bibitem{Vos:IJCFD2011}
P.~E.~J. Vos, C.~Eskilsson, A.~Bolis, S.~Chun, R.~M. Kirby, and S.~J. Sherwin.
\newblock {A generic framework for time-stepping partial differential equations
  (PDEs): general linear methods, object-oriented implementation and
  application to fluid problems}.
\newblock {\em International Journal of Computational Fluid Dynamics},
  25(3):107--125, Mar. 2011.

\end{thebibliography}

\end{document}